\input phyzzx
\catcode`@=11 
\def\space@ver#1{\let\@sf=\empty \ifmmode #1\else \ifhmode
   \edef\@sf{\spacefactor=\the\spacefactor}\unskip${}#1$\relax\fi\fi}
\def\attach#1{\space@ver{\strut^{\mkern 2mu #1} }\@sf\ }
\newtoks\foottokens
\newbox\leftpage \newdimen\fullhsize \newdimen\hstitle
\newdimen\hsbody
\newif\ifreduce  \reducefalse
\def\almostshipout#1{\if L\lr \count2=1
      \global\setbox\leftpage=#1 \global\let\lr=R
  \else \count2=2
    \shipout\vbox{\special{dvitops: landscape}
      \hbox to\fullhsize{\box\leftpage\hfil#1}} \global\let\lr=L\fi}
\def\smallsize{\relax
\font\eightrm=cmr8 \font\eightbf=cmbx8 \font\eighti=cmmi8
\font\eightsy=cmsy8 \font\eightsl=cmsl8 \font\eightit=cmti8
\font\eightt=cmtt8
\def\eightpoint{\relax
\textfont0=\eightrm  \scriptfont0=\sixrm
\scriptscriptfont0=\sixrm
\def\rm{\fam0 \eightrm \f@ntkey=0}\relax
\textfont1=\eighti  \scriptfont1=\sixi
\scriptscriptfont1=\sixi
\def\oldstyle{\fam1 \eighti \f@ntkey=1}\relax
\textfont2=\eightsy  \scriptfont2=\sixsy
\scriptscriptfont2=\sixsy
\textfont3=\tenex  \scriptfont3=\tenex
\scriptscriptfont3=\tenex
\def\it{\fam\itfam \eightit \f@ntkey=4 }\textfont\itfam=\eightit
\def\sl{\fam\slfam \eightsl \f@ntkey=5 }\textfont\slfam=\eightsl
\def\bf{\fam\bffam \eightbf \f@ntkey=6 }\textfont\bffam=\eightbf
\scriptfont\bffam=\sixbf   \scriptscriptfont\bffam=\sixbf
\def\tt{\fam\ttfam \eightt \f@ntkey=7 }
\def\caps{\fam\cpfam \tencp \f@ntkey=8 }\textfont\cpfam=\tencp
\setbox\strutbox=\hbox{\vrule height 7.35pt depth 3.02pt width\z@}
\samef@nt}
\def\Eightpoint{\eightpoint \relax
  \ifsingl@\subspaces@t2:5;\else\subspaces@t3:5;\fi
  \ifdoubl@ \multiply\baselineskip by 5
            \divide\baselineskip by 4\fi }
\parindent=16.67pt
\itemsize=25pt
\thinmuskip=2.5mu
\medmuskip=3.33mu plus 1.67mu minus 3.33mu
\thickmuskip=4.17mu plus 4.17mu
\def\thinspace{\kern .13889em }
\def\negthinspace{\kern-.13889em }
\def\enspace{\kern.416667em }
\def\enskip{\hskip.416667em\relax}
\def\quad{\hskip.83333em\relax}
\def\qquad{\hskip1.66667em\relax}
\def\crr{\cropen{8.3333pt}}
\foottokens={\Eightpoint\singlespace}
\def\papersize{\SIZE\OFFSET\skip\footins=\bigskipamount}
\def\SIZE{\hsize=11.8truecm\vsize=17.5truecm}
\def\OFFSET{\voffset=-1.3truecm\hoffset=  .14truecm}
\message{STANDARD CERN-PREPRINT FORMAT}
\def\attach##1{\space@ver{\strut^{\mkern 1.6667mu ##1} }\@sf\ }
\def\PH@SR@V{\doubl@true\baselineskip=20.08pt plus .1667pt minus
.0833pt
             \parskip = 2.5pt plus 1.6667pt minus .8333pt }
\def\author##1{\vskip\frontpageskip\titlestyle{\tencp ##1}\nobreak}
\def\address##1{\par\kern 4.16667pt\titlestyle{\tenpoint\it ##1}}
\def\andaddress{\par\kern 4.16667pt \centerline{\sl and} \address}
\def\abstract{\vskip2\frontpageskip\centerline{\tenrm Abstract}
              \vskip\headskip }
\def\cases##1{\left\{\,\vcenter{\Tenpoint\m@th
    \ialign{$####\hfil$&\quad####\hfil\crcr##1\crcr}}\right.}
\def\matrix##1{\,\vcenter{\Tenpoint\m@th
    \ialign{\hfil$####$\hfil&&\quad\hfil$####$\hfil\crcr
      \mathstrut\crcr\noalign{\kern-\baselineskip}
     ##1\crcr\mathstrut\crcr\noalign{\kern-\baselineskip}}}\,}
\Tenpoint
}
\def\Smallsize{\smallsize\reducetrue
\let\lr=L
\hstitle=8truein\hsbody=4.75truein\fullhsize=24.6truecm\hsize=\hsbody
\output={
  \almostshipout{\leftline{\vbox{\makeheadline
  \pagebody\makefootline}}}\advancepageno
     }
\special{dvitops: landscape}
\def\makeheadline{
\iffrontpage\line{\the\headline}
             \else\vskip .0truecm\line{\the\headline}\vskip .5truecm
\fi}
\def\makefootline{\iffrontpage\vskip  0.truecm\line{\the\footline}
               \vskip -.15truecm\line{\the\date\hfil}
              \else\line{\the\footline}\fi}
\paperheadline={
\iffrontpage\hfil
               \else
               \tenrm\hss $-$\ \folio\ $-$\hss\fi    }
\paperstyle}
%
%
%
%
%
%
%
%
%
\newcount\referencecount     \referencecount=0
\newif\ifreferenceopen       \newwrite\referencewrite
\newtoks\rw@toks
\def\NPrefmark#1{\attach{\scriptscriptstyle [ #1 ] }}
\let\PRrefmark=\attach
\def\refmark#1{\relax\ifPhysRev\PRrefmark{#1}\else\NPrefmark{#1}\fi}
\def\refend{\refmark{\number\referencecount}}
\newcount\lastrefsbegincount \lastrefsbegincount=0
\def\refsend{\refmark{\count255=\referencecount
   \advance\count255 by-\lastrefsbegincount
   \ifcase\count255 \number\referencecount
   \or \number\lastrefsbegincount,\number\referencecount
   \else \number\lastrefsbegincount-\number\referencecount \fi}}
\def\refch@ck{\chardef\rw@write=\referencewrite
   \ifreferenceopen \else \referenceopentrue
   \immediate\openout\referencewrite=referenc.texauxil \fi}
%
{\catcode`\^^M=\active 
  \gdef\obeyendofline{\catcode`\^^M\active \let^^M\ }}%
%
{\catcode`\^^M=\active 
  \gdef\ignoreendofline{\catcode`\^^M=5}}
{\obeyendofline\gdef\rw@start#1{\def\t@st{#1} \ifx\t@st\blankend%
\endgroup \@sf \relax \else \ifx\t@st\bl@nkend \endgroup \@sf \relax%
\else \rw@begin#1
\backtotext
\fi \fi } }
{\obeyendofline\gdef\rw@begin#1
{\def\n@xt{#1}\rw@toks={#1}\relax%
\rw@next}}
\def\blankend{}
{\obeylines\gdef\bl@nkend{
}}
\newif\iffirstrefline  \firstreflinetrue
\def\rwr@teswitch{\ifx\n@xt\blankend \let\n@xt=\rw@begin %
 \else\iffirstrefline \global\firstreflinefalse%
\immediate\write\rw@write{\noexpand\obeyendofline \the\rw@toks}%
\let\n@xt=\rw@begin%
      \else\ifx\n@xt\rw@@d \def\n@xt{\immediate\write\rw@write{%
        \noexpand\ignoreendofline}\endgroup \@sf}%
             \else \immediate\write\rw@write{\the\rw@toks}%
             \let\n@xt=\rw@begin\fi\fi \fi}
\def\rw@next{\rwr@teswitch\n@xt}
\def\rw@@d{\backtotext} \let\rw@end=\relax
\let\backtotext=\relax

\newdimen\refindent     \refindent=30pt
\def\refitem#1{\par \hangafter=0 \hangindent=\refindent
\Textindent{#1}}
\def\REFNUM#1{\space@ver{}\refch@ck \firstreflinetrue%
 \global\advance\referencecount by 1 \xdef#1{\the\referencecount}}
\def\refnum#1{\space@ver{}\refch@ck \firstreflinetrue%
 \global\advance\referencecount by 1
\xdef#1{\the\referencecount}\refend}

\def\REF#1{\REFNUM#1%
 \immediate\write\referencewrite{%
 \noexpand\refitem{#1.}}%
\begingroup\obeyendofline\rw@start}
\def\ref{\refnum\?%
 \immediate\write\referencewrite{\noexpand\refitem{\?.}}%
\begingroup\obeyendofline\rw@start}
\def\Ref#1{\refnum#1%
 \immediate\write\referencewrite{\noexpand\refitem{#1.}}%
\begingroup\obeyendofline\rw@start}
\def\REFS#1{\REFNUM#1\global\lastrefsbegincount=\referencecount
\immediate\write\referencewrite{\noexpand\refitem{#1.}}%
\begingroup\obeyendofline\rw@start}
\newif\ifmref  
\newif\iffref  
\def\xrefsend{\xrefmark{\count255=\referencecount
\advance\count255 by-\lastrefsbegincount
\ifcase\count255 \number\referencecount
\or \number\lastrefsbegincount,\number\referencecount
\else \number\lastrefsbegincount-\number\referencecount \fi}}
\def\xrefsdub{\xrefmark{\count255=\referencecount
\advance\count255 by-\lastrefsbegincount
\ifcase\count255 \number\referencecount
\or \number\lastrefsbegincount,\number\referencecount
\else \number\lastrefsbegincount,\number\referencecount \fi}}
\def\xREFNUM#1{\space@ver{}\refch@ck\firstreflinetrue%
\global\advance\referencecount by 1
\xdef#1{\xrefend}}
\def\xrefend{\xrefmark{\number\referencecount}}
\def\xrefmark#1{[{#1}]}
\def\xRef#1{\xREFNUM#1\immediate\write\referencewrite%
{\noexpand\refitem{#1 }}\begingroup\obeyendofline\rw@start}%
\def\xREFS#1{\xREFNUM#1\global\lastrefsbegincount=\referencecount%
\immediate\write\referencewrite{\noexpand\refitem{#1 }}%
\begingroup\obeyendofline\rw@start}
\def\rrr#1#2{\relax\ifmref{\iffref\xREFS#1{#2}%
\else\xRef#1{#2}\fi}\else\xRef#1{#2}\xrefend\fi}
\def\multref#1#2{\mreftrue\freftrue{#1}%
\freffalse{#2}\mreffalse\xrefsend}
\referencecount=0
\def\refbreak{\hfil\penalty200\hfilneg}
\def\paperstyle{\papers}
\paperstyle   
%
%
%
\def\slacpub{\afterassignment\slacp@b\toks@}
\def\slacp@b{\edef\n@xt{\Pubnum={\the\toks@}}\n@xt}
\let\pubnum=\slacpub
\expandafter\ifx\csname eightrm\endcsname\relax
    \let\eightrm=\ninerm \let\eightbf=\ninebf \fi

\font\seventeencp=cmcsc10 scaled\magstep3

\newif\ifCONF \CONFfalse
\newif\ifBREAK \BREAKfalse
\newif\ifsectionskip \sectionskiptrue

%
%
%
%
\def\NuclPhysProc{
\let\lr=L
\hstitle=8truein\hsbody=4.75truein\fullhsize=21.5truecm\hsize=\hsbody
\hstitle=8truein\hsbody=4.75truein\fullhsize=20.7truecm\hsize=\hsbody
\output={
  \almostshipout{\leftline{\vbox{\makeheadline
  \pagebody\makefootline}}}\advancepageno
     }
\def\papersize{\SIZE\OFFSET\skip\footins=\bigskipamount}
\def\SIZE{\hsize=10.0truecm\vsize=27.0truecm}
\def\OFFSET{\voffset=-1.4truecm\hoffset=-2.40truecm}
\message{NUCLEAR PHYSICS PROCEEDINGS FORMAT}
\def\makeheadline{
\iffrontpage\line{\the\headline}
             \else\vskip .0truecm\line{\the\headline}\vskip .5truecm
\fi}
\def\makefootline{\iffrontpage\vskip  0.truecm\line{\the\footline}
               \vskip -.15truecm\line{\the\date\hfil}
              \else\line{\the\footline}\fi}
\paperheadline={\hfil}
\paperstyle}
%
%
%
%

%
%
%
%

%
%
%
%
\def\ReprintVolume{\smallsize
\def\papersize{\hsize=18.0truecm\vsize=23.1truecm\voffset -.73truecm
    \hoffset -.65truecm\skip\footins=\bigskipamount
    \normaldisplayskip= 20pt plus 5pt minus 10pt}
\message{REPRINT VOLUME FORMAT}
\paperstyle\baselineskip=.425truecm\parskip=0truecm
\def\makeheadline{
\iffrontpage\line{\the\headline}
             \else\vskip .0truecm\line{\the\headline}\vskip .5truecm
\fi}
\def\makefootline{\iffrontpage\vskip  0.truecm\line{\the\footline}
               \vskip -.15truecm\line{\the\date\hfil}
              \else\line{\the\footline}\fi}
\paperheadline={
\iffrontpage\hfil
               \else
               \tenrm\hss $-$\ \folio\ $-$\hss\fi    }
\def\sectionfont{\bf}    }
%
%
%
%
\def\SIZE{\hsize=15.73truecm\vsize=23.11truecm}
\def\OFFSET{\voffset=0.0truecm\hoffset=0.truecm}
\message{DEFAULT FORMAT}
\def\papersize{\SIZE\OFFSET\skip\footins=\bigskipamount
\normaldisplayskip= 35pt plus 3pt minus 7pt}
\Pubnum={\rm \the\pubnum }
\def\title#1{\vskip\frontpageskip\vskip .50truein
     \titlestyle{\seventeencp #1} \vskip\headskip\vskip\frontpageskip
     \vskip .2truein}
\def\author#1{\vskip .27truein\titlestyle{#1}\nobreak}

\def\p@bblock{\begingroup \tabskip=\hsize minus \hsize
   \baselineskip=1.5\ht\strutbox \topspace+2\baselineskip
   \halign to\hsize{\strut ##\hfil\tabskip=0pt\crcr
  \the \Pubnum\cr}\endgroup}
\def\makefootline{\iffrontpage\vskip .27truein\line{\the\footline}
                 \vskip -.1truein
              \else\line{\the\footline}\fi}
\paperfootline={\iffrontpage\message{FOOTLINE}
\hfil\else\hfil\fi}

\def\abstract{\vskip2\frontpageskip\centerline{\twelvebf Abstract}
              \vskip\headskip }

\paperheadline={
\iffrontpage\hfil
               \else
               \twelverm\hss $-$\ \folio\ $-$\hss\fi}
%
%
\def\nup#1({\refbreak\ Nucl.\ Phys.\ $\underline {B#1}$\ (}
\def\plt#1({\refbreak\ Phys.\ Lett.\ $\underline  {#1}$\ (}
\def\cmp#1({\refbreak\ Commun.\ Math.\ Phys.\ $\underline  {#1}$\ (}
\def\prp#1({\refbreak\ Physics\ Reports\ $\underline  {#1}$\ (}
\def\prl#1({\refbreak\ Phys.\ Rev.\ Lett.\ $\underline  {#1}$\ (}
\def\prv#1({\refbreak\ Phys.\ Rev. $\underline  {D#1}$\ (}
\def\und#1({            $\underline  {#1}$\ (}
%
%

\def\rB{\hfil\penalty1000\hfilneg}
%
%
\hyphenation{sym-met-ric anti-sym-me-tric re-pa-ra-me-tri-za-tion
Lo-rentz-ian a-no-ma-ly di-men-sio-nal two-di-men-sio-nal}
%
%
%
%

\def\coeff#1#2{{\textstyle { #1 \over #2}}\displaystyle}
\def\boxit#1{\vbox{\hrule\hbox{\vrule\kern3pt
\vbox{\kern3pt#1\kern3pt}\kern3pt\vrule}\hrule}}
\message{ by V.K, W.L and A.S}
\catcode`@=12
\paperstyle
\input tables

\def\chi {\X}

\paperstyle

\def\mod{{\rm ~mod~}}

\def\half{\coeff12}

\def\Zbf{{\bf Z}}

\def\X{{\cal X}}

\def\ishiket#1{{\ket{#1}}\rangle}
\def\ishibra#1{\langle{\bra{#1}}}
\catcode`@=11
\def\ninef@nts{\relax
    \textfont0=\ninerm          \scriptfont0=\sixrm
      \scriptscriptfont0=\sixrm
    \textfont1=\ninei           \scriptfont1=\sixi
      \scriptscriptfont1=\sixi
    \textfont2=\ninesy          \scriptfont2=\sixsy
      \scriptscriptfont2=\sixsy
    \textfont3=\tenex          \scriptfont3=\tenex
      \scriptscriptfont3=\tenex
    \textfont\itfam=\nineit     \scriptfont\itfam=\seveni  
\sevenit
    \textfont\slfam=\ninesl     \scriptfont\slfam=\sixrm 
\sevensl
    \textfont\bffam=\ninebf     \scriptfont\bffam=\sixbf
      \scriptscriptfont\bffam=\sixbf
    \textfont\ttfam=\tentt
    \textfont\cpfam=\tencp }
\def\ninepoint{\ninef@nts \samef@nt \b@gheight=9pt \setstr@t }
\newif\ifnin@  \nin@false
\def\Tenpoint{\tenpoint\twelv@false\nin@false\spaces@t}
\def\Twelvepoint{\twelvepoint\twelv@true\nin@false\spaces@t}
\def\Ninepoint{\ninepoint\twelv@false\nin@true\spaces@t}
\def\spaces@t{\rel@x
      \iftwelv@ \ifsingl@\subspaces@t3:4;\else\subspaces@t1:1;\fi
       \else \ifsingl@\subspaces@t3:5;\else\subspaces@t4:5;\fi \fi
      \ifdoubl@ \multiply\baselineskip by 5
         \divide\baselineskip by 4 \fi
       \ifnin@ \ifsingl@\subspaces@t3:8;\else\subspaces@t4:7;\fi \fi
}
\def\Vfootnote#1{\insert\footins\bgroup
   \interlinepenalty=\interfootnotelinepenalty \floatingpenalty=20000
   \singl@true\doubl@false \iftwelv@ \Tenpoint
   \else \Ninepoint \fi
   \splittopskip=\ht\strutbox \boxmaxdepth=\dp\strutbox
   \leftskip=\footindent \rightskip=\z@skip
   \parindent=0.5\footindent \parfillskip=0pt plus 1fil
   \spaceskip=\z@skip \xspaceskip=\z@skip \footnotespecial
   \Textindent{#1}\footstrut\futurelet\next\fo@t}

\def\small#1{\vskip .3truecm\footnoterule\nobreak
\Ninepoint\parindent=2pc\sl
\hang #1 \vskip .3truecm\nobreak\footnoterule}

\def\small#1{}

  

%
\def\PSSA{\rrr\PSSA{
G.~Pradisi, A.~Sagnotti and Ya.S.~Stanev, Phys Lett B354 (1995) 279
[arXiv:hep-th/9503207].}}
\def\PSSB{\rrr\PSSB{
G.~Pradisi, A.~Sagnotti and Ya.S.~Stanev,
Phys. Lett B356 (1995) 230
[arXiv:hep-th/9506014].}}
\def\FPS{\rrr\FPS{
D.~Fioravanti, G.~Pradisi and A.~Sagnotti,
Phys.\ Lett.\ B {\bf 321}, 349 (1994)\rB
[arXiv:hep-th/9311183].}}
\def\PSSC{\rrr\PSSC{
G.~Pradisi, A.~Sagnotti and Ya.S.~Stanev, Phys Lett B354 (1995) 279;
 Phys. Lett. B381 (1996) 97
[arXiv:hep-th/9603097].}}
\def\BPPZ{\rrr\BPPZ{
R.~Behrend, P.~Pearce, V.~Petkova and J-B.~Zuber,
Phys.\ Lett.\ B {\bf 444}, 163 (1998)
[arXiv:hep-th/9809097]; Nucl.\ Phys.\ B {\bf 579}, 707 (2000)
[arXiv:hep-th/9908036].}}

\def\DiF{\rrr\DiF{
P.~Di~Francesco and J.-B Zuber, Nucl.~Phys. {\bf B338} (1990)
602.}}

\def\YaS{\rrr\YaS{
A.N. Schellekens and Ya.S. Stanev. JHEP {\bf 0112}, 012 (2001)\rB
[arXiv:hep-th/0108035].}}

\def\Cardy{\rrr\Cardy{
J. Cardy, Nucl. Phys. B324 (1989) 581.}}

\def\HSS{\rrr\HSS{L.~Huiszoon, B.~Schellekens and N.~Sousa,
Nucl. Phys. {\bf B575} (2000) 401.}}

\def\HSSK{\rrr\HSSK{L.~Huiszoon, B.~Schellekens and N.~Sousa,
Phys. Lett. {\bf B470} (1999) 95.}}

\def\HuiszoonGE{\rrr\HuiszoonGE{
L.~R.~Huiszoon and A.~N.~Schellekens,
Nucl.\ Phys.\ B {\bf 584}, 705 (2000)
[arXiv:hep-th/0004100].}}

\def\FHSSW{\rrr\FHSSW{J.~Fuchs, L.~Huiszoon, B.~Schellekens, C.~Schweigert,
J.~Walcher,
Phys. Lett. {\bf B495} (2000) 427, [arXiv:hep-th/0007174].}}

\def\FSA{\rrr\FSA{J.~Fuchs, C.~Schweigert, Phys. Lett.
{\bf B490} (2000) 163, [arXiv:hep-th/0006181].}}

\def\FSB{\rrr\FSB{L. Birke, J. Fuchs, C. Schweigert, Adv. Theor. Math. Phys.
{\bf 3} (1999) 671, [arXiv:hep-th/9905038].}}

\def\FuchsXN{\rrr\FuchsXN{
J.~Fuchs and C.~Schweigert,
Nucl.\ Phys.\ B {\bf 568}, 543 (2000)
[arXiv:hep-th/9908025]; Nucl.\ Phys.\ B {\bf 558}, 419 (1999)
[arXiv:hep-th/9902132].}}

\def\FuchsCM{\rrr\FuchsCM{
J.~Fuchs, I.~Runkel and C.~Schweigert,
[arXiv:hep-th/0204148].}}

\def\SchM{\rrr\SchM{A.N.~Schellekens,
\cmp 153 (1993)  159.
}}

\def\bantay{\rrr\bantay{P. Bantay,
 Phys. Lett B 394 (1997) 87.}}

\def\CoSu{\rrr\CoSu{F.~Bais and
P.~Bouwknegt,
\nup279 (1987) 561; \rB A.N.~Schellekens and N.P.~Warner,
\prv34 (1986) 3092.}}

\def\GannonKI{\rrr\GannonKI{
T.~Gannon,
Nucl.\ Phys.\ B {\bf 627}, 506 (2002)
[arXiv:hep-th/0106105].}}

\def\FuchsSQ{\rrr\FuchsSQ{
J.~Fuchs, A.~N.~Schellekens and C.~Schweigert,
Nucl.\ Phys.\ B {\bf 437}, 667 (1995)
[arXiv:hep-th/9410010].}}

\def\FuchsKT{\rrr\FuchsKT{
J.~Fuchs and C.~Schweigert,
Phys.\ Lett.\ B {\bf 414}, 251 (1997)
[arXiv:hep-th/9708141].}}

\def\PetkovaYJ{\rrr\PetkovaYJ{
V.~B.~Petkova and J.~B.~Zuber,
``Boundary conditions in charge conjugate sl(N) WZW theories,''
arXiv:hep-th/0201239.}}

\def\NUNO{\rrr\NUNO{
A.N. Schellekens and N.~Sousa, Int. J. Mod. Phys. A16 (2001) 3659.}}

\def\GannonJM{\rrr\GannonJM{
T.~Gannon, P.~Ruelle and M.~A.~Walton,
Commun.\ Math.\ Phys.\  {\bf 179}, 121 (1996)
[arXiv:hep-th/9503141].}}

\def\FSS{\rrr\FSS{
J.~Fuchs, A.~N.~Schellekens and C.~Schweigert,
Nucl.\ Phys.\ B {\bf 473}, 323 (1996)
[arXiv:hep-th/9601078].}}

\def\SchellekensWW{\rrr\SchellekensWW{
A.~N.~Schellekens,
Int.\ J.\ Mod.\ Phys.\ A {\bf 14}, 1283 (1999)
[arXiv:math.qa/9806162].}}

\def\FSSD{\rrr\FSSD{
J.~Fuchs, B.~Schellekens and C.~Schweigert,
Commun.\ Math.\ Phys.\  {\bf 180}, 39 (1996)
[arXiv:hep-th/9506135].}}

\def\MoSb{\rrr\MoSb{
G.~Moore   and N.~Seiberg,
\nup313 (1988) 16.}}
\def\ScYe{\rrr\ScYe{A.N.~Schellekens and S.~Yankielowicz,
\nup334 (1990) 67.}}

\def\planar{\rrr\planar{G. Pradisi, A. Sagnotti and Ya.S. Stanev, {\em Planar  
duality in $SU(2)$ WZW models}, Phys. Lett. B354 (1995) 279  
[arXiv:hep-th/9503207].}}

\def\ZuberIA{\rrr\ZuberIA{
J.~B.~Zuber,
{\em CFT, BCFT, ADE and all that}, lectures at Bariloche 2000,
[arXiv:hep-th/0006151].}}

\def\Pasq{\rrr\Pasq{
V.~Pasquier, J. Phys. A. 20 (1987) 5707.}}

\def\CIZ {\rrr\CIZ {
A.~Cappelli, C.~Itzykson     and J.-B.~Zuber,
\nup280 (1987)  445;\rB \cmp113 (1987) 1.}}
\def\Gep {\rrr\Gep {
D.~Gepner,
\nup287 (1987)  111.}}
%

\def\PetkovaFW{\rrr\PetkovaFW{
V.~B.~Petkova and J.~B.~Zuber,
Nucl.\ Phys.\ B {\bf 463}, 161 (1996)\rB
[arXiv:hep-th/9510175].}}

\pubnum={{}}
\rightline{NIKHEF/02-009}
\rightline{hep-th/0210014}
\rightline{September 2002}
\date{September 2002}
\pubtype{CRAP}
\titlepage
\message{TITLE}

\title{\fourteenbf Orientation matters for NIMreps }
\author{N. Sousa and A.N. Schellekens\foot{t58@nikhef.nl}}
\line{\hfil
 {\it NIKHEF}
 \hfil}
\line{\hfil \it NIKHEF, P.O. Box 41882, 1009 DB Amsterdam,
The Netherlands  \hfil}
\bigskip

\abstract \noindent

The problem of finding boundary states in CFT, often rephrased in
terms of ``NIMreps"  of the fusion algebra, has a natural extension
to CFT on non-orientable surfaces. This provides extra information that
turns out to be quite useful to give the proper interpretation to a
NIMrep. We illustrate this with several examples. This includes a
rather detailed discussion of the interesting case of the
simple current extension of
$A_2$ level 9, which
is already known to have a rich structure. This structure can be disentangled
completely using orientation information. In particular we find
here and in other cases examples of {\it diagonal} modular invariants that
do not admit a NIMrep, suggesting that there does not exist a corresponding
CFT.
We obtain the
complete set of NIMreps
(plus Moebius and Klein bottle coefficients) for many exceptional
modular invariants of WZW models, and find an explanation for the occurrence
of more than one NIMrep in certain cases.
We also (re)consider the
underlying formalism, emphasizing the distinction between oriented and
unoriented string annulus amplitudes, and the origin of
orientation-dependent degeneracy matrices
in the latter.

\baselineskip= 15.0pt plus .2pt minus .1pt

\chapter{Introduction}

Boundary states in conformal field theory (CFT) have been studied for at
least two rather different reasons. First of all one may need these
states in a particular physical problem in which boundaries of the Riemann
surface are important, such as open string theory or
applications of statistical mechanics. But on the other hand it has also
been suggested that the mere (non)-existence of (a complete set \PSSC\ of) boundaries
may tell us something about the existence of the CFT itself.

Completeness of boundaries in usually rephrased \PSSC,\BPPZ\ in terms of an
equation involving the annulus coefficients $A_{ia}^{~~b}$, which are
non-negative integers appearing in the decomposition of the open string
spectrum (more precise definitions will be given below). These integer  
matrices must form a non-negative integer matrix
representation (NIMrep) of the fusion algebra:
$$ A_{ia}^{~~b} A_{jb}^{~~c}= N_{ij}^{~~k} A_{ka}^{~~c} \eqn\AANA $$
A general solution
found by Cardy \Cardy\ is to choose $A$ equal to the fusion coefficients.
The issue of finding
non-trivial NIMreps was considered shortly thereafter in \DiF\ in the
context of RSOS models. These authors also gave some non-trivial solutions,
in particular for $SU(3)$ WZW models.
The relation to completeness of boundaries was understood in \PSSC, after
which many papers appeared giving solutions in various cases with various
degrees of explicitness, see \eg\ \BPPZ\ and
\multref\PSSA{\PSSB\FuchsKT\FuchsXN\FSA\FSB\NUNO\FHSSW\GannonKI\ZuberIA\PetkovaYJ}.

The link of a given NIMrep
to closed CFT is made by simultaneously diagonalizing the integer
matrices $A$ in terms of irreducible representations of fusion algebra
and matching the representations that appear with the C-diagonal terms in
the torus partition function matrix $Z_{ij}$
(by ``C-diagonal" we mean the coefficients
$Z_{ii^c}$).

The issue of existence of a CFT
 usually arises in situations where
the existence of a modular invariant partition
function (MIPF) is the only hint of a possible CFT.
This occurs frequently in the case of rational CFT's,
where such partition functions are determined by a non-negative integer
matrix that commutes with the modular transformation matrices, a
simple algebraic
problem that can be solved by a variety of methods, including brute force.
It is well known that the existence of a MIPF is
necessary, but not
sufficient for the existence of a CFT.
Unfortunately the necessary additional
checks are considerably harder to carry out, since they
involve fusing and braiding matrices that are available in only a few
cases (for recent progress and an extensive list of references
see \FuchsCM).
It has been suggested
that
the existence of a complete set of boundaries might be used as an additional
consistency check. This point was in particular emphasized
recently by Gannon \GannonKI.

The implication of this point of view is that a
CFT that cannot accommodate
a complete set of boundaries is inconsistent even if one is only interested in
closed Riemann surfaces. By the same logic, a symmetric\foot{We mean here
a CFT with a symmetric modular invariant partition function.}
CFT that cannot exist
on unorientable surfaces, i.e. Riemann surfaces with crosscaps, should then
be inconsistent, even if one is only interested in orientable surfaces.

The advantage of using crosscap consistency is that there is a very
simple consistency check, namely
$$ \sum _m S_{im} K_m = 0\ , \hbox{if $i$ is ${\underline{\rm not}}$ an  
Ishibashi label}\eqn\KBSum$$
where $K_m$ are the Klein bottle coefficients, which must satisfy the
conditions $K_m = Z_{mm} \mod 2$ and $|K_m| \leq Z_{mm}$, where $Z_{mn}$ is
the integer matrix defining the torus partition function.

There are, however, two caveats.
First of all, non-symmetric modular invariants manifestly
do not allow non-orientable surfaces, since an orientation reversal
interchanges the left and right Hilbert spaces. Nevertheless such partition
functions may well correspond to sensible CFT's. It is not obvious that
asymmetry of the modular invariant is the {\it only} possible
obstruction to orientation reversal. In other words, there might exist
examples of symmetric CFT's allowing boundaries but no crosscaps.
Secondly, the Klein bottle sum rule \KBSum\ can only be used successfully with
some additional restrictions on the allowable Klein bottle coefficients, since
the integrality constraints usually allow too many solutions. These additional
restrictions are unfortunately only conjectures, and hence
on less firm ground than the integrality
constraints. One of them, the
``Klein bottle constraint", that requires the signs of the
non-vanishing coefficients $K_i$ to be preserved in fusion:
$$ K_i K_j K_k > 0 {\hbox{~~~~if~~~}} N_{ij}^{~~k} \not = 0 \ , \eqn\KBC $$
is known to be
violated in some simple current examples \HSS. Therefore either there is
something wrong with those examples, or
a more precise formulation
of the Klein bottle constraint is needed (we consider the latter possibility
to be the most likely one).
The other restriction is the trace
formula \YaS
$$ \sum_a A^i_{aa} = {S^{i\ell}\over S_{0\ell}} Y^{\ell}_{~00} K_{\ell} \ ,
\eqn\kleintrace$$
to which no counter examples are known, but which is
still only a conjecture. In this paper we will consider
the extra information from non-orientable surfaces always in conjunction
with boundary information, and we will see that violations of  \KBSum\ usually
give the right hint regarding the existence of boundaries and presumably the
existence of the CFT. We investigated the validity of the Klein bottle
constraint as well as the trace formula \kleintrace\ in many examples,
and found additional counter examples to the former, but additional
support for the conjecture \kleintrace.

Another way in which orientation information can be used is to search for
Moebius and Klein bottle amplitudes belonging to an already given NIMrep.
This can be done by solving a set of ``polynomial equations" given in
\YaS. In this case only the first caveat applies: if the polynomial equations
have no solution, then either the modular invariant is not sensible or
it does not admit an orientifold. Without further information this cannot
be decided unambiguously, but the examples strongly suggest a definite answer.

The orientation information is encoded in terms of a set of integer
data that extends the notion of a NIMrep.
This is most naturally done in two stages, first to an
``S-NIMrep" (symmetrized NIM-rep) and then to a
``U-NIMrep" (unoriented NIMrep), by adding Moebius and Klein bottle
coefficients. A given NIMrep can admit several S-NIMreps (or none at all),
which in their turn may have any number of U-NIMreps.
For example, the charge conjugation
invariant has a well-known NIMrep (the Cardy solution), which extends to at  
least one U-NIMrep, but often to more than one \planar\HSSK, obtained using
simple currents.
For all simple current modifications\foot{Note that
simple currents are used in two distinct ways here: to change the Klein
bottle projection for a given modular invariant, and to change the
modular invariant itself.}of charge conjugation modular
invariant a class of
U-NIMreps has been determined in \FHSSW\ (this includes all U-NIMreps
that exist for generic simple current invariants, but there might
exist additional exceptional solutions). In \PSSB\ the U-NIMreps
were presented for all the $A_1$ automorphism invariants, including
the exceptional ``$E_7$" invariant at level 16.
Another interesting class are
the diagonal invariants. A general prescription for finding their NIMreps
was given in \FSB. This involves the construction of a ``charge conjugation
orbifold", which does not necessarily exist. If the orbifold CFT
{\it does} exist, the NIMrep
is constructed from it using simple currents. Then, using the results of
\FHSSW, one may also construct U-NIMreps for the diagonal invariant.
The charge conjugation
orbifold theory was constructed explicitly for WZW-models (recently,
the NIMreps for SU(N) diagonal invariants were obtained by different
methods in \PetkovaYJ, while the case $N=3$ was already considered in \DiF;
these authors did not consider U-NIMreps). Finally, it is straightforward to
extend the known NIMreps for the exceptional invariants of $A_1$ \BPPZ\
to U-NIMreps. This summarizes what is presently known
about U-NIMreps.

As remarked above, the existence of (S,U)-NIMreps may be used as
a guiding principle to the (non)-existence of a CFT.
In \GannonKI\ several examples were given of MIPF without a corresponding
NIMrep.
In these examples, the MIPF is an extension of the chiral algebra.
It was already known on other grounds that this extension was inconsistent,
and hence the non-existence of the complete set of boundaries just gives
an additional confirmation. Here we will present some examples that
are automorphism invariants, for which no simple consistency checks
are available, other than modular invariance itself. We will find examples that
do not admit a complete set of boundaries, and examples that do not admit
crosscap coefficients, even though a complete set of boundaries does exist.  
Perhaps surprisingly, among
the examples for which a complete set of boundaries does not exist
are diagonal invariants (\ie\ the Cardy case modified by charge
conjugation). This presumably implies
 that the charge conjugation orbifold theory
does not exist.

Apart from existence of a NIMrep, one may also worry about uniqueness
of a NIMrep. At least one example is known \DiF\GannonKI\ where a given modular
invariant appears to have more than one NIMrep. We find many more,
and show how the ambiguity is resolved by computing in addition
to the annulus coefficients also the Moebius and Klein bottle
coefficients. Multiple NIMreps can be expected if there are degeneracies
in the spectrum, \ie\ if $Z_{ii^c} > 1$ for some $i$. There are exceptions:
if the degeneracies are due to simple currents, or if the degeneracy
is absorbed into the extended characters (rather than leading to several
Virasoro-degenerate extended characters) we see no reason to expect multiple
NIMreps.

This paper is consists of two parts: a discussion of the formalism
for general modular invariants, and a set of examples illustrating
several special features. In the next chapter we discuss
the formalism, with particular emphasis on the distinction between
annuli for oriented and unoriented strings. We believe this clarifies
some points that have at best remained implicit in the existing
literature. This discussion is presented for the general case,
allowing for Ishibashi state multiplicities larger than 1, which
usually occur for modular invariants of extension type.
Chapters 3-7 contain various classes of examples, illustrating in
different ways what can be learned about NIM-reps by paying attention
to orientation issues. The examples in section 3 concern automorphisms
of simple current extended WZW models. Here we discover an example
of a diagonal invariant for which no NIMrep exists. In chapter 4 we
discuss this in more generality, and find
some additional examples of this kind. In section 5 we consider
automorphisms of $c=1$ orbifolds.
Since our results suggest that not all the allowed automorphisms of
extended WZW-models yield sensible CFT's, it is natural to inspect also
pure WZW automorphisms. These were
completely enumerated in \GannonJM. We examine some of these
MIPF's in section 6, and we find no inconsistencies. Finally in section
7 we consider exceptional extensions of WZW models, including
cases with multiplicities larger than 1 and extensions by higher spin
(\ie\ higher than 1) currents. The results for the spin 1
extensions can be understood
remarkably well from the point of view of the extended theory, whereas
the results for higher spin extensions confirm the consistency of the new
CFT's obtained after the extension.

We will not present the explicit NIMreps here in order to save space, but
all data are available from the authors on request.
The examples
where obtained by solving \AANA\ on a computer, using the fact
that all matrix elements are bounded from above
by quantum dimension \GannonKI.
Apart from \AANA\ we also
imposed the condition that the matrix representation should be diagonalizable
in terms of the correct one-dimensional fusion representations, with
multiplicity $\Zbf_{ii^c}$. Although this is a finite search in principle,
in practice a straightforward search is impossible in
essentially all cases. However, using a variety of
methods  -- which will not be
explained here --
we were able to do
an exhaustive search in most cases of interest.

\chapter{Orientation issues for NIM-reps}

Here want to discuss two issues that have remained unsatisfactory
in the literature so far:
\item\dash Two different expressions for ``annulus coefficients" are
in use, one of the form ``$S{\cal B}{\cal B}^*$" and the other form ``$S{\cal  
B}{\cal B}$"
\item\dash For a given CFT there are sometimes several choices for the complete
set of boundary coefficients.

Here ``$S$" stands for the
modular transformation matrix and ``${\cal B}$" for the boundary
coefficients (for reasons explained later we  use the notation ${\cal B}$  
instead of $B$ here).
The indices of these quantities are as follows.
A complete basis for the boundaries in a CFT are the Ishibashi states,
labelled by a pair of chiral representations $(i,i^c)$. The MIPF
determines how often each Ishibashi state actually appears in a CFT.
The completeness hypothesis (recently proved in \FuchsCM)
states that the number of
boundaries is equal to the number of Ishibashi states.
Given
a MIPF, parametrized by a non-negative integral matrix $Z_{ij}$,
the number of boundaries should thus be given by $\sum_i Z_{ii^c}$.
Since an Ishibashi state may appear more than once we need an
additional degeneracy label $\alpha$, so that the relevant Ishibashi
labels are
$(i,\alpha), \alpha=1,\ldots, Z_{ii^c}$. We denote the Ishibashi states
as $\ishiket{i,\alpha}$.
The coupling
of such a state to boundary $a$ is given by a boundary coefficient
${\cal B}_{(i,\alpha)a}$, and the completeness condition states that this should be a
square, invertible matrix. The boundary coefficients appear in the
expansion of boundary states in terms of Ishibashi states,
$\ishiket{{\cal B}_a} = \sum_{i,\alpha}{\cal B}_{(i,\alpha)a} \ishiket{i,\alpha}$

Regarding the first point above, some authors do seem to understand that
the expression $S{\cal B}{\cal B}^*$ is to be used for oriented strings, and  
$S{\cal B}{\cal B}$
for unoriented ones, but this is rarely stated, and these expressions
are often presented without justification. The former expression is the
one obtained most straightforwardly.
The usual derivation of the
open string partition function is to transform to the closed string channel,
where the amplitude can be written as a closed string exchange between
two branes.
This amplitude
can be evaluated easily in the transverse (closed string tree) channel
by taking matrix elements of the closed string propagator $U$ between
boundary states:  $\ishibra{B_b} U \ishiket {B_a}$.
 This describes closed strings propagating from boundary
$a$ to boundary $b$.  Transforming this expression
back to the open string loop channel
 results in the following expression
for the annulus coefficients
$$ A^{~~b}_{ia}=\sum_j\sum_{\alpha} S_{ij}
 {\cal B}_{(j,\alpha)a}{\cal B}_{(j,\alpha)b}^* \eqn\ABB $$
In general the resulting expression is not symmetric in the two indices $a$ and $b$,
indicating that this amplitude describes oriented strings. Obviously a symmetric
expression is obtained by dropping the complex conjugation, but that step
would require some justification.

Regarding the second point, the paradox is that on general
grounds one would expect there to exist a unique set of boundary coefficients $B$,
given the CFT and the MIPF. This is because they are the complete set
of irreducible (and hence one-dimensional) representations
of a ``classifying algebra" \FuchsKT\PSSC\Pasq.
This is an abelian algebra of the
form\foot{Our normalization differs from \FuchsKT, in order to
obtain a simpler form for the annuli.}
$$  B_{(i,\alpha)a} B_{(j,\beta)a} =
\sum_{k,\gamma} X_{(i,\alpha)(j,\beta)}^{~~~~~~~~~~(k,\gamma)} S_{0a}  
B_{(k,\gamma)a} \eqn\classalg $$
whose coefficients $X$ can be
expressed in terms of CFT data like OPE coefficients and fusing matrices.
Although these coefficients are rarely available in explicit form, they
are completely fixed by the algebra.
Nevertheless it was found that in many cases -- where \classalg\ is not  
available and only
integrality conditions were solved --
a given bulk
CFT may have more than one set of boundary coefficients ${\cal B}$, giving rise
to different
annulus coefficients when the form $S{\cal B}{\cal B}$ is used. In all these
cases, the different annuli are related to each other by changing
the coefficients ${\cal B}_{(i,\alpha)a}$ by a $(i,\alpha)$-dependent phase.
This changes the annulus coefficients in the $S{\cal B}{\cal B}$
form, but it does not change the $S{\cal B}{\cal B}^*$ annuli. Obviously such phases
do not respect the classifying algebra\rlap,\foot{Alternatively, one may
define orientation-dependent generalizations of the
classifying algebra that incorporate the sign changes, as was done in \YaS.  
However, there will always
be one special choice with structure coefficients derived from the CFT data in
the canonical way.}and hence at most one of the different sets ${\cal B}$ can  
be equal to the
unique solution $B$ to \classalg. We  use the symbol $B$ to denote the
solution to \classalg, and ${\cal B}$ to denote any complete set of boundary  
coefficients that
satisfy the annulus integrality conditions. In principle, one can compute  
orientable annuli
(NIMreps)
$A_{ia}^{~~b}$ for each choice of ${\cal B}$, and yet another one by replacing  
${\cal B}$ by the
solution $B$ of \classalg. If indeed the coefficients ${\cal B}$ and $B$  
differ only by
phases (or unitary matrices) on the degeneracy spaces, all these NIMreps are  
identical, and
we might as well write  $A_{ia}^{~~b}$ in terms of $B$:
$$ A^{~~b}_{ia}=\sum_j\sum_{\alpha} S_{ij}
 {B}_{(j,\alpha)a}{B}_{(j,\alpha)b}^* \eqn\ABBtwo $$

Recent work on orientifold planes in group manifolds provides some useful
geometrical insight into the solution of the two problems stated above. The  
different boundaries
$\ishiket{B_a}$  correspond to D-branes at different positions. The coefficients
$B_{(j,\alpha)a}$ specify the couplings of these branes to closed strings.
If one adds orientifold
planes, there are no changes to the positions of the D-branes, but they are
identified by the orientifold reflection. Hence one would expect the couplings
$B_{(j,\alpha)a}$ to remain unchanged, in agreement with the classifying algebra
argument given above. Nevertheless there is an important change in the system:  
without
O-planes, a stack of $N_a$ D-branes gives rise to a $U(N_a)$ Chan-Paton gauge group.
In the presence of an O-plane, self-identified branes give rise to $SO$ or  
$Sp$ groups,
whereas pairwise identified branes give $U$ groups.

In string theory, oriented open strings occur when one considers branes that are
not space-time filling in an oriented closed string theory, for example the  
type-II superstring.
They give rise to $U(N)$ Chan-Paton groups. The $U(N)$ gauge bosons come from  
the spectrum of
open strings with both endpoints attached to the same brane.
When the branes are space-time filling tadpole cancellation
(or RR-charge cancellation) requires the the addition of O-planes. In this  
case, $SO$, $Sp$ and $U$
can occur, and the $U(N)$ gauge bosons come from open string with their  
endpoints attached
to {\it different} branes, identified by the orientifold map.
This suggests that in order to describe such a system one should consider a
different partition function, which in the closed string channel correspond
to propagation of strings from a brane to the orientifold reflection of a brane.

An orientifold projection involves an operator ${\Omega}$
that reverses the orientation of the worldsheet. In a geometric description
${\Omega}$ does not only invert the worldsheet orientation, but it may
also act non-trivially on the target space. This operator has some
realization in CFT (denoted by the same symbol $\Omega$) which in any case
interchanges the left and right Hilbert space, and which may have a non-trivial
action on the representations, which must square to 1. On Ishibashi states  
this yields\foot{The matrix $\Omega$ introduced here is a generalization
of the signs $\epsilon$ defined in \PSSB.}
$$ \Omega \ishiket{i,\alpha} = \sum_{\beta} \Omega^{i}_{\alpha,\beta}  
\ishiket{i^c,\beta} $$
Of course the action of $\Omega$ must be defined on all closed string states, but we
only need it on the Ishibashi states here. The condition that $\Omega$ is a  
reflection
implies
$$\sum_{\beta}\Omega^i_{\alpha\beta}\Omega^{i^c}_{\beta\gamma} =  
\delta_{\alpha\gamma} \eqn\Square$$
Furthermore $\Omega$ must be unitary, i.e
$$ \sum_{\beta}\Omega^i_{\alpha\beta}(\Omega^i_{\gamma\beta})^* =  
\delta_{\alpha\gamma} \ .\eqn\Unit$$
This implies
$$ \Omega^i = (\Omega^{i^c})^{\dagger} \eqn\OmegaDagger$$
The reasoning in the foregoing paragraph suggests that the quantity of interest is
$\ishibra{B_b} U \Omega \ishiket {B_a}$, transformed to the open string channel.
The computation results in
$$ A^{\Omega}_{iab} =\sum_j\sum_{\alpha,\beta} S_{ij}
 B_{(j,\alpha)a}\Omega^{j}_{\alpha\beta}B_{(j^c,\beta)b}^* \ .\eqn\Aomega $$
If this is the right expression, it should be symmetric. To show that it is, one
may replace the left hand side by its complex conjugate (since the right hand side
should obviously be real), use \OmegaDagger, change the summation from $j$ to  
$j^c$, and
finally use $S_{ij}=(S_{ij^c})^*$.

In order to write \Aomega\ in manifestly symmetric form we need a relation of the
form
$$ (B_{(j^c,\alpha)a})^* = \sum_{\beta} C^j_{\alpha\beta} B_{(j,\beta),a} \ ,  
\eqn\CPTB $$
where $C^j$ must be unitary, and is uniquely determined if we know $B$.
Such a relation should always hold as a consequence of CPT invariance, since
it relates the emission of a closed string state $(j,j^c)$ to the absorption
of its charge conjugate. Conjugating twice we find
$\sum_{\beta}(C^j_{\alpha\beta})^*C^{j^c}_{\beta\gamma}=\delta_{\alpha\gamma}$.
This relation may be used to show that
$$ A^{\Omega}_{iab} =\sum_j\sum_{\alpha,\beta} S_{ij}
 B_{(j,\alpha)a}[\Omega^{j}C^j]_{\alpha\beta} B_{(j,\beta)b} \ .\eqn\AomegaTwo $$
is indeed symmetric, if $A^{\Omega}_{iab}$ is real. The latter relation
may be written as
$$ A^{\Omega}_{iab} =\sum_j\sum_{\alpha,\beta} S_{ij}
 B_{(j,\alpha)a}g^j_{\alpha\beta} B_{(j,\beta)b} \ .\eqn\AomegaThree $$
Conversely, if $A^{\Omega}_{iab}$ is symmetric and $B$ is invertible
(as it is in the case of a complete set of boundaries), then $g^j_{\alpha\beta}$
must be symmetric. Such a ``degeneracy matrix"
was first introduced in \FHSSW\ for a rather different purpose, and further
discussed in \YaS. Here we even encounter it in the absence of
degeneracies. Note that one can of course take a square root of $g^j$ and
absorb it into the definition of $B_{(j,\alpha)a}$. This the yields the quantity
previously denoted as {\cal B}.
This was implicitly
or explicitly done in several previous papers,
\eg\ \FHSSW\ and \YaS, and leads to
orientation-dependent boundary coefficients. Both formalisms
are of course completely equivalent if one is only interested
in partition functions.

Note that in most cases we do not know $g^j$ (or $\Omega^j$)
and $B_{(j,\alpha)}$ separately, but we only know the integer data
$A^{i~~b}_{~a}$ or $A^{\Omega}_{iab}$. Obviously, from the former we
cannot determine $\Omega$, and we can only determine $B_{(j,\alpha)}$ up
to Ishibashi-dependent phases (or unitary matrices in case of degeneracies).
From $A^{\Omega}_{iab}$ we can determine $B_{(j,\alpha)}$ up to signs (or
orthogonal matrices in case of degeneracies), provided we know $\Omega$.
Furthermore, since the coefficients $B_{(j,\alpha)}$ as well as the
matrices $C^j$ are assumed to be independent of the orientifold choice,
we can determine relative signs of two choices of $\Omega$.
In the examples discussed so far two choices of $\Omega$ (or $g^j$) always had
{\it relative} eigenvalues $\pm 1$, but that need not be the case in general.
If one
knows the classifying algebra coefficients one can obviously
determine $B$ completely, and then also $\Omega$ and $C$.

The allowed choices of $\Omega$ are determined by the requirement
that it must be a symmetry of the CFT. A necessary condition is that
the coefficients $\Omega^j$ must respect
 the OPE's of the bulk CFT, and hence presumably the fusion rules.
For automorphism invariants (\ie\ no degeneracies) an obvious solution
to that condition is
$\Omega^j=1$, but there might be further constraints in the full CFT.
Obviously these matrices will have to satisfy the appropriate
generalization (to allow for degeneracies) of the crosscap constraint
of \FPS\ and \PSSB, but we will not pursue that here, because we are
only considering constraints that do not require knowledge of fusing matrices
and OPE coefficients.

A related issue is that of boundary conjugation. The boundary
conjugation matrix is defined as
$$ C^B_{ab}=A^{\Omega}_{0ab} \ .$$
The matrix $A^{\Omega}_{0ab}$ must be an involution in order to get
meaningful Chan-Paton groups in string theory.
Since it is an involution, we can define $a^c$ as the boundary
conjugate to $a$, satisfying $C^B_{aa^c}=1$.
The boundary conjugation matrix may be
used to raise and
lower indices. Then we may define
$$ A_{ia}^{\Omega~~b} = \sum_c A^{\Omega}_{iac} A^{\Omega}_{0cb} \eqn\NimOm$$
For a complete set of boundaries this quantity satisfies eqn \AANA.
If the latter
have a unique solution (for a given set of Ishibashi states),
it follows that all quantities $A_{ia}^{\Omega~~b}$ are
identical. Conversely, one can say that all quantities $A^{\Omega}_{iab}$ are  
different
symmetrizations of $A^{~~b}_{ia}$. We will refer to such a symmetrization as an
S-NIMrep.

Given a NIMrep, one can always write the oriented annulus
coefficient $A_{ia}^{~~b}$ in the form \ABBtwo, since this just amounts
to diagonalizing a set of commuting normal\foot{A normal matrix
is a square matrix that commutes with its adjoint. This is automatically
true if $A_{ia}^{~~b} = A_{i^cb}^{~~~a}$ and $A_{ia}^{~~b}$ is real, properties
that should be treated as part of the definition of a NIMrep. It is not hard
to show that the matrices $A_{ia}^{\Omega~~b}$ satisfy this.}matrices  
\ZuberIA. We will now show that given an
S-NIMrep, one can write $A^{\Omega}_{iac}$ in the form $\AomegaThree$.
Note that\foot{We are assuming here for simplicity that the NIMrep
$A_{ia}^{\Omega~~b^c}$ is unique, so we may drop the superscript $\Omega$.
If it is not unique, not only $A_{ia}^{~~b^c}$ but also $B$ get an
additional label $\Omega$, but the rest of the discussion is unaffected.}
$$ A^{\Omega}_{iab}  =  A_{ia}^{~~b^c} = A^{\Omega}_{iba} = A_{ib}^{~~a^c}$$
so that
$$ A_{ia^c}^{~~b^c} = A^{\Omega}_{iba^c} = A_{ib}^{~~a}\ . $$
Contracting this with a matrix $S_{im}$ we get
$$ \sum_{\alpha} B_{(m,\alpha)a^c}(B_{(m,\alpha)b^c})^*
  = \sum_{\alpha} B_{(m,\alpha)b}(B_{(m,\alpha)a})^* \ ,$$
where $B_{(m,\alpha)a}$ are the set of matrices in terms of which
\ABBtwo\  holds.
Completeness implies that the matrix $B_{(m,\alpha)a^c}$ has a right
inverse $X$,
$$ \sum_a  B_{(m,\alpha)a^c}  
X_{a^c(n,\gamma)}=\delta_{mn}\delta_{\alpha\gamma}\ . $$ This leads to the
following expression
$$ \sum_{\alpha} \delta_{mn}\delta_{\alpha\gamma}(B_{(m,\alpha)b^c})^*
  = \sum_{\alpha}  B_{(m,\alpha)b}\sum_a(B_{(m,\alpha)a})^*X_{a^c(n,\gamma)} \ . $$
which implies
$$ \eqalign{ (B_{(m,\gamma)b^c})^*
  &= \sum_{\alpha}  B_{(m,\alpha)b}\sum_a(B_{(m,\alpha)a})^*X_{a^c(m,\gamma)}\cr
&= \sum_{\alpha}  B_{(m,\alpha)b}V^{m}_{\alpha\gamma} \ .\cr}$$
Iterating this we see that $V^m$ must be a unitary matrix. Hence we get the
desired answer,
$$ A^{\Omega}_{iab} = \sum S_{im}B_{(m,\alpha)a})
V^{m}_{\alpha\gamma} B_{(m,\gamma)b} $$
Since the left hand side is symmetric, $V^m$ must be a symmetric matrix.
It can be written as $\Omega^m C^m$ to extract $\Omega^m$, but nothing
guarantees that the resulting $\Omega$ is indeed a symmetry. Indeed there may well
exist symmetrizations that have no interpretation in terms of orientifold maps.

We now have two ways of arriving at an S-NIMrep, a.) by
computing $\ishibra{B_b} U \Omega \ishiket{B_a}$ and
transforming
to the open string loop channel, and b.) by starting from
the oriented amplitude $\ishibra{B_b} U  \ishiket{B_a}$, and contracting it
 with
$C^B_{ab}$. Comparing the two expressions and using completeness
of boundaries, we arrive at the following expression for the action
of $\Omega$ on a boundary state
$$ \Omega \ishiket{B_a} = C^B_{ab}  \ishiket{B_b} \ . \eqn\omegaonboundary $$

An S-NIMrep is only interesting as a first ingredient in a complete
set of annuli, Moebius and Klein amplitudes. We will refer to such a
set of data as an U-NIMrep (with ``U" for unoriented). This problem can also
be phrased (almost) entirely in terms of integers \YaS. The full set of
conditions is then that the following quantities must exist
\item\dash NIMrep: a set of non-negative integer matrices
$A_{ia}^{~~b}=A_{i^cb}^{~~a}$
satisfying \AANA
\item\dash S-NIMrep: In addition, an involution $C^B_{ab}$ such that
$A_{iab} \equiv A_{ia}^{~~c}C^B_{cb}$ is symmetric
\item\dash U-NIMrep: In addition to the two foregoing requirements,
a set of Moebius coefficients $M_{ia}$ and Klein bottle coefficients $K_i$
such that $\half(A_{iaa}+M_{ia})$ and $\half(Z_{ii}+K_i)$  are non-negative  
integers, and the following polynomial equations are satisfied
$$ \eqalign{\sum_b A_{ia}^{~~b}M_{jb}&=\sum_l Y_{ij}^{~~l} M_{la}  \cr
  \sum_{a,b} C^B_{ab} M_{ia}M_{jb} &= \sum_l Y^l_{~ij}K_l\ . \cr} \eqn\polynom$$

\noindent The derivation of \polynom\ in \YaS\ uses the following
formulas for $M_{\ell a}$ and $K_{\ell}$.
$$ K_{\ell} = \sum_{m,\alpha} S_{\ell m}
\Gamma_{m\alpha}g^{m}_{\alpha\beta}\Gamma_{m\beta} \eqn\Kform $$
$$ M_{\ell a} = \sum_{m,\alpha} P_{\ell m}
\Gamma_{m\alpha}g^{m}_{\alpha\beta}B_{(m\beta)a} \eqn\Mform $$
In the computation one uses the completeness relation
$$ \sum_a S_{0m }B_{(m,\alpha)a}B_{(\ell,\beta)a}^* =
\delta_{m\ell}\delta_{\alpha\beta}\ ,\eqn\complone $$
which follows from $A_{0a}^{~~b}=\delta_a^b$ plus the
completeness assumption
that $B$ is an invertible matrix. In an analogous way one derives
from $A_{0ab}=C^B_{ab}$:
$$ \sum_{a,b} S_{0m }B_{(m,\alpha)a}C^B_{ab}B_{(\ell,\beta)b} =
\delta_{m\ell}(g^m)^{-1}_{\alpha\beta}\ ,\eqn\compltwo $$
which implies
$$ B_{(m,\alpha)a}^*=\sum_{b,\beta}C^B_{ab}g^m_{\alpha\beta}B_{(m,\beta)b}
\eqn\Bstar $$

The formulas for $M_{\ell a}$ and $K_{\ell}$ follow from
the closed string amplitudes
$\ishibra{B_a} U \Omega \ishiket{\Gamma}$ and
$\ishibra{\Gamma} U  \Omega \ishiket{\Gamma}$, rather than the corresponding
expressions without $\Omega$.
Actually, this should not matter, since
we expect $\ishiket{\Gamma}$ to be an eigenstate of $\Omega$: a crosscap
corresponds to an orientifold plane, formed by the fixed points of
$\Omega$.
 This would still seem to allow two
possibilities:
$$ \Omega \ishiket{\Gamma} = \pm  \ishiket{\Gamma} \ ,$$
but there are several reasons to believe that the $+$ sign
is the correct one. First of all $\Omega$ acts on self-conjugate
boundaries only with a $+$ sign, which might suggest that the
same should be true for O-planes; secondly,
if this relation holds with a $-$ sign, one finds   
$M_{ia}\hat\chi_i=-M_{ia^c}\hat\chi_i$ (where $\hat \chi$ are
the usual Moebius characters).
This
is obviously inconsistent for self-conjugate boundaries, which must
have a non-vanishing $M_{0a}$. Hence in theories with at least one
self-conjugate boundary, $ \ishiket{\Gamma}$ {\it must} be an $\Omega$
eigenstate with eigenvalue $+1$. This includes most CFT's but leaves
open the possibility of a rare exception.
Finally, if different expressions were used for the annulus, Moebius
and Klein bottle amplitudes, tadpole factorization for
the genus 1 open string amplitudes would not work.
While none of these arguments are totally convincing, it seems
nevertheless reasonable to assume that
$\Omega \ishiket{\Gamma} =  \ishiket{\Gamma} $, which implies
$ M_{ia}\hat\chi_i=M_{ia^c}\hat\chi_i $.
In CFT's with degenerate Virasoro
characters the implication for the Moebius coefficients is
ambiguous, but we can be more precise.

The relation $\Omega \ishiket{\Gamma} =  \ishiket{\Gamma} $
implies a relation for the basis coefficients $\Gamma_{m,\alpha}$
appearing in the expansion
$$ \ishiket{\Gamma}=\sum_{m,\alpha} \Gamma_{(m,\alpha)}\ishiket{m,\alpha} \ , $$
namely
$$ \Gamma_{m^c,\beta}=\sum_{\alpha}\Gamma_{m,\alpha}\Omega^m_{\alpha\beta}
\eqn\OmegaGamma $$
Furthermore we will assume that the coefficients $\Gamma$ satisfy the same
CPT relation \CPTB\ as the coefficients $B$, \ie\
$$ (\Gamma_{(j^c,\alpha)})^* = \sum_{\beta} C^j_{\alpha\beta}  
\Gamma_{(j,\beta)} \ , \eqn\CPTC$$
which is plausible since the action is entirely on the space of Ishibashi
states. Then we can derive
$$ \eqalign{
M_{\ell^c a^c} &= \sum_{m,\alpha} P_{\ell^c m}
\Gamma_{m,\alpha}g^{m}_{\alpha\beta}B_{(m\beta)a^c} \cr
 &= \sum_{m,\alpha,\beta,\gamma} P_{\ell m}
  C^{m^c}_{\alpha\beta}  \Gamma_{m,\gamma}\Omega^m_{\gamma\beta}
B_{(m\alpha)a} \cr
 &= \sum_{m,\alpha,\beta,\gamma} P_{\ell m}
  \Gamma_{m,\gamma}\Omega^m_{\gamma\beta}C^{m}_{\beta\alpha}
B_{(m\alpha)a} = M_{\ell a} \cr
}
\eqn\MMC$$
where we used \Bstar, reality of $M_{\ell,a}$, \CPTC, \OmegaGamma\ and
finally the relation $C^{m^c} = (C^m)^T$ derived just after \CPTB.
This relation is an additional constraint that must hold  if
$\Omega$ is actually a symmetry. In the next chapter we will
encounter solutions to \polynom\ that violates this constraint, and
must therefore be rejected.

On might think that
the polynomial equations \polynom\ are not sufficient, just
as the NIMrep condition \AANA\ is not sufficient to relate a NIMrep
to a modular invariant.
We must also require that in
the loop channel only valid Ishibashi states propagate.
In other words, the proper channel transformation applied to
the Moebius and Klein bottle coefficients that solve these equations
should give zero on non-Ishibashi labels.
In fact, more is required: a
set of boundary coefficients $B_{(i,\alpha)a}$ and crosscap
coefficients $\Gamma_{(i,\alpha)}$ that reproduce
the integers. We will now show that this already
follows from \polynom, \ie\ that
\Mform\ and \Kform\ not only imply \polynom, but that the converse
is also true.

Suppose the set of integers $M_{ia}$ satisfies the first equation.
We can always write them as $M_{ia}=\sum_m P_{im} X_{ma}$ since
$P$ is invertible. Plugging this into the first equation and using
\ABBtwo, plus several inversions of $P$ and $S$ we get
$$ X_{\ell a}=\sum_{\alpha}B_{(\ell,\alpha)a}
\left[\sum_b B_{(\ell,\alpha)b} X_{\ell b}\right]
\equiv \sum_{\alpha} B_{(\ell,\alpha)a}C_{(\ell,\alpha)}\ .\eqn\Xdef$$
Substituting the resulting expression for $M_{ia}$ into the second equation
we find that any set of Klein bottle coefficients satisfying it must be
of the form
$$ K_{i} = \sum_{m,\alpha,\beta}S_{im} C_{m,\alpha} (g^m)^{-1}_{\alpha\beta}
C_{m,\beta}\ , \eqn\KleinFormula $$
where we used \compltwo. Defining
 $C_{m,\alpha}=\sum_{\beta} g_{\alpha\beta}\Gamma_{m,\beta}$ then gives
us both $K_{i}$ and $M_{ia}$ in the required form. Note that \Xdef\ does
not determine the crosscap coefficients outside the space spanned by the
Ishibashi states. However, \KleinFormula\ shows that there is
no room for additional components, so that $C_{m,\alpha}$ must vanish
on non-Ishibashi labels.
Hence no
further conditions are required.

Note that in both cases there are far more equations than variables.
In practice the first set usually reduces to a number of independent equations
slightly smaller than the number of Moebius coefficients
(or equal, in which case the only solution is $M_{ia}=K_i=0$).
 Since the equations
are homogeneous, the open string integrality conditions are needed to cut
the space of solutions down to a finite set. The second set of equations
produces a unique set of coefficients $K_{\ell}$ for any solution
of the first. These coefficients are then subject to the closed sector
integrality conditions.

One may expect the symmetry $\Omega$ to be closely related to the
choice of Klein bottle projection $K_i$, which determines the
projection in the closed string sector.
The link between the two is
provided by a relation postulated in \YaS:
$$ \sum_{a,\alpha\beta}S_{0j}B_{(j,\alpha)a} g^j_{\alpha\beta} B_{(j,\beta)a}
= Y^j_{~00}K_{j}\ , \eqn\YaSconj $$
conjectured to hold {\it without} summation on $j$. Here we restored the
dependence on $g$, that was absorbed into the definition of $B$ in \YaS.
Even though we do not know $g^j$ itself, it is clear that if this relation
holds, and if we limit ourselves to automorphism invariants and to
real CFT's\rlap,\foot{Note that $Y^j_{~00}$ is equal to the
Frobenius-Schur indicator \bantay\HSSK, which vanishes for complex fields.} any
sign change of $\Omega^j$ implies a sign change of $K_j$ for the same
value of $j$. However, this does not necessarily
imply that $\Omega^i$ is equal to $K^i$.

A closely related issue is the so-called ``Klein bottle constraint".
This condition would imply that the signs of the Klein bottle
 are preserved by the fusion rules. This is obviously
true if $\Omega^i=K^i$, since $\Omega$ is a symmetry. However,
there are cases where the Klein bottle constraint is violated in
otherwise consistent simple current U-NIMreps \HSS. In those
examples either  $\Omega^i\not=K^i$, or $\Omega$ is not a symmetry.
Since we cannot determine $\Omega$ without further information, as
discussed earlier, this cannot be decided at present.
One important check can be made, however. The examples admit (at least)
two choices for $K^i$, both violating the Klein bottle constraint.
The sign flips relating these choices imply, via \YaSconj, corresponding
sign flips in $\Omega^i$, which must themselves respect the fusion
rules if each choice of $\Omega^i$ does. We have checked that this is
indeed true.

\chapter{Examples I: Automorphisms of Extended WZW models}

In the following chapters we consider a variety of modular
invariant partition functions of CFT's, for which we compute
all the NIMreps that satisfy $A^{i~~b}_{~a}=A^{i^c~a}_{~b}$,
all the symmetrizations of each NIMrep, and
all the Moebius and Klein bottle coefficients that satisfy the
polynomial equations \polynom, together
with the usual mod-2 conditions.

Additional constraints that may
be imposed are
\point The Klein bottle constraint \KBC.
\point The orientifold condition \MMC.
\point The trace formula \kleintrace.
\point Positivity of the Klein bottle coefficient of the vacuum.
\point Reality of the crosscap coefficients\rlap.\foot{This is a conjecture
due to A. Sagnotti.}

We have seen examples violating any of these conditions, but Nr. 3
was only violated if Nr. 4 (which is clearly necessary in applications
to string theory) was violated as well. In those cases it turns
out that \kleintrace\ holds
with a $-$ sign, and furthermore there was a second solution that
satisfies \kleintrace\ and has the signs of all Klein bottle coefficients
reversed (note that the polynomial equations allow a sign flip
of all Moebius coefficients, but not in general a sign flip of all
Klein bottle coefficients).
We consider any violation of conditions Nr 1,2,3 and 4 as
signs of an inconsistency, but we will mention such solutions whenever
they occur.

The number of solutions we present is after removing all
equivalences (\ie\ the overall Moebius sign choice, boundary permutations
that respect the NIMrep and relate different S-NIMreps, and boundary
permutations that respect an S-NIMrep and relate different sets of
Moebius  coefficients).

The examples we consider in this chapter
 occur in CFT's obtained by extending the
chiral algebra of a WZW-model by simple currents. We emphasize that
we work in the extended CFT and hence we only consider boundaries that
respect the extended symmetries. Nevertheless we will get information
about broken boundaries from other modular invariants of the extended
theory.
The cases we consider have one fixed point representation, and we
apply the method of fixed point resolution of \FSS\ to obtain the modular
transformation matrix of the extended theory.

We will be interested in automorphism modular invariants of the extended
theory. Usually the fixed point resolution leads to a
CFT that has one or more non-trivial fusion rule automorphisms in which
the resolved fixed points are interchanged. Here we consider three cases
in which there are additional automorphisms interchanging resolved fixed
points with other representations. This phenomenon was first
observed in the so-called ``$E_7$"-type modular invariant \CIZ\
of $A_1$ level 16 \Gep,
which is the first in a short series including also $A_2$ level 9 \MoSb\ and
$A_4$ level 5 \ScYe.

\section{$A_1$ level 16}

Consider first as a warm-up example
$A_1$ level 16.
The extended theory has six primaries,
which we label as (0), (1), (2), (3), ($4^+$) and ($4^-$), where the
integers indicate the smallest $SU(2)$-spin in the ground state
representation. The spin 4 representation is a fixed point of the simple
current, and hence is split into two degenerate representations. In
addition the representation (1) has the same conformal weight
as ($4^+$) and ($4^-$), up to integers. All representations are
self-conjugate.

This extended algebra has six  modular
invariants, corresponding to the six permutations of
$(1)$,
$(4^+)$ and $(4^-)$
$$ \eqalign{
   a:\quad& (0)^2+(1)^2+(2)^2+(3)^2+(4^+)^2+(4^-)^2 \cr
c:\quad& (0)^2+(1)^2+(2)^2+(3)^2+(4^+)\times (4^-)+(4^-)\times (4^+) \cr
b^+:\quad& (0)^2+(1)\times (4^+)+(2)^2+(3)^2+(4^+)\times (1)+(4^-)^2 \cr
b^-:\quad& (0)^2+(1)\times (4^-)+(2)^2+(3)^2+(4^-)\times (1)+(4^+)^2 \cr
d^+:\quad& (0)^2+(1)\times (4^+)+(2)^2+(3)^2+(4^-)\times (1)+(4^+)\times (4^-)  
\cr
d^-:\quad& (0)^2+(1)\times (4^-)+(2)^2+(3)^2+(4^+)\times (1)+(4^-)\times (4^+)  
\cr} $$
When we write these partition functions
in terms of $SU(2)$ characters (which do not distinguish
$(4^+)$ from $(4^-)$) on obtains respectively the invariants
of types\break  $(D,D,E,E,E,E)$ of $A_1$.

The number of Ishibashi labels is 6,4,4,4,3,3 respectively, and we found
precisely one complete set of boundaries in all six cases (for the unextended
theory, the NIMreps were presented in the second paper of \BPPZ; the U-NIMrep
of the invariants ``b" in the extended theory were obtained in \PSSB.)
All these sets are symmetric and have no other symmetrizations.
Furthermore in all cases except $d^+$ and $d^-$ a U-NIMrep exists.
In this
case we know that all these boundaries are indeed physically meaningful.
The cases $(a,c)$, $(b^+,d^+)$ and $(b^-,d^-)$ are pairwise related
to each other by a left conjugation (\ie\ a $(4^+)$, $(4^-)$ interchange),
whereas $(a,c)$, $(b^+,d^-)$ and $(b^-,d^+)$ are pairwise related by
a right conjugation. These conjugations are like T-dualities, which from
the point of view of the unextended $A_1$ theory (the orbifold theory of
this conjugation)
interchange the automorphism type of the boundary.
Hence the 6 boundaries
of the $a$-invariant plus the 4 boundaries of the $c$-invariant together
form the 10 boundaries expected for the $D$-type invariant of $A_1$ level
16; the 4 boundaries of the $d$-invariant plus the 3 boundaries of
the $e$-invariant form the 7 boundaries of the
$E$-type invariant of $A_1$ level
16.

In the extended theory, orientation does not add much information, since
all boundary coefficients were expected to exist, and since furthermore the
cases $d^+$ and $d^-$ are asymmetric invariants that do not admit
an orientifold projection.

\section{$A_2$ level 9}

Now consider another case with similar features, namely
$A_2$ level 9. The extended theory now has 8 primaries, which we
will label as follows
$$ \eqalign {
(0):\quad& (0,0)\quad\quad
(1):\quad (0,3)\quad\quad
(2):\quad (3,0)\cr
(3):\quad& (1,1)\quad\quad
(4):\quad (4,4)\quad\quad
(5):\quad (2,2)\cr
(6^i):\quad& (3,3),\quad i=1,2,3\cr}$$
Here $(a,b)$ are $A_2$ Dynkin labels of one of the ground state
representations. The representations (1) and (2) are conjugate to each other,
and all others are self-conjugate.

The extended theory has a total of 48 distinct modular invariants,
obtained by combining charge conjugation with the 24 permutations
of (3), $(6^1)$, $(6^2)$ and $(6^3)$. These modular invariants
are related to the following four distinct ones of $A_2$ level 9\foot{
Here ``D" stands for ``type D" as in ADE, and not for ``diagonal". Note
that we have included a charge conjugation in the definitions, so that
D corresponds to the C-diagonal or Cardy case, and DC to the diagonal
invariant.}
$$\eqalign{
D:\quad &[0]^2 +[1][2] +[2][1]+[3]^2+[4]^2+[5]^2 + 3 [6]^2 \cr
DC:\quad &[0]^2 +[1]^2 +[2]^2+[3]^2+[4]^2+[5]^2 + 3 [6]^2 \cr
E:\quad &[0]^2 +[1][2] +[2][1]+[3][6]+[6][3]+[4]^2+[5]^2 + 2 [6]^2\ \cr
EC:\quad &[0]^2 +[1]^2 +[2]^2+[3][6]+[6][3]+[4]^2+[5]^2 + 2 [6]^2 , \cr} $$
where $[i]$, $i\not=6$ stands for the full simple current orbit of
the orbit representatives listed above, and $[6]$ is
the fixed point. These modular invariants
admit respectively 21, 15, 17 and 11 Ishibashi states.

In the following table we show the number of NIMreps (of various kinds)
that exist for these 48 invariants. The results of this table
are based on an exhaustive search, \ie\ the set of solutions is
complete.

\vskip .7truecm
\begintable
Nr. | Permutation | Proj. | Ish. | NIMreps | S-NIMreps| U-NIMreps \cr
1 | $<>$       | D |    9        |        1        |     1          |      1   \nr
2 |$<a,b>$    | D   |    7        |     0          |    0            |     0 \nr
3 |$<3,b>$    | E   |    7        |     0          |    0            |     0 \nr
4 |$<1,2>$     | DC  |    7        |     0          |    0            |     0 \nr
5 |$<1,2><a,b>$ | DC    |    5        |     1          |    2            |     
$1+1(^*)$      \nr
6 |$<1,2><3,b>$ | EC    |    5        |     1          |    2            |     
$1+1(^*)$      \nr
7 |$ <3,a><b,c>$ | E  |    5        |     1          |    2            |     1  \nr
8 |$<3,a,b,c>$   |E |    5        |     1          |       2          |  1     \nr
9 |$<1,2><3,a><b,c>$ | EC   |    3        |     1          |    1            |  
  0 \nr
10 |  $<1,2><3,a,b,c>$ | EC   |    3        |     1          |       1          
 |  0     \nr
11 |$<a,b,c>$  | D  |    6        |     $2(^{**})$          |       0           
|  0     \nr
12 |$<3,a,b>$  | E  |    6        |     $2(^{**})$          |       0           
|  0     \nr
13 |$<1,2><a,b,c>$ |DC   |    4        |     0          |       0          |   
0   \nr
14 | $<1,2><3,a,b>$ |EC   |    4        |     0          |       0          |   
0     \endtable
\leftline{(*) Inconsistent solution to the polynomial equations.}
\leftline{(**) Asymmetric NIMrep, one solution is the transpose of the other.}

In this table, $a$, $b$ and $c$ denote the three resolved fixed points.
The solutions are essentially the same if $a$, $b$ or $c$ are replaced by $3$,  
but we
represent these cases separately because they have a different interpretation.
Furthermore cases 7,8 and 9,10 have pairwise the same solutions, because
boundary CFT is only sensitive to the diagonal and C-diagonal partition
function, which is identical in these cases (note that when solving
\polynom\ we do not take into account the left-right symmetry of the
modular invariant).
In cases 5 and 6 the
notation is as follows: the unique NIMrep allows two symmetrizations.
For each of these symmetrizations there is one set of Moebius
coefficients that satisfy the polynomial equations \polynom\ (there are
40 independent equations for the 45 Moebius coefficients). The
Moebius
coefficients for the second S-NIMrep (indicated with a (*) in the table)
should be rejected, however. They give rise to complex crosscap coefficients and
to Klein bottle coefficients that violate the Klein bottle constraint.
More importantly, however, they violate the condition $M_{ia}=M_{i^ca^c}$,  
indicating that there is no orientifold
symmetry that underlies these coefficients.

As was the case for $A_1$ we may expect these NIMreps to correspond to those
of the unextended theory. Furthermore the symmetry breaking boundaries may be
expected to correspond to two different modular invariants of the extended theory,
obtained by applying the cyclic permutations $<a,b,c>$ and $<a,c,b>$ to the  
``symmetric"
partition function. This leads to the following identifications of
partition functions of the unextended theory with triplets of extended
theory partition functions

$$\eqalign{
D:\quad &<>; \ <a,b,c>; \ <a,c,b> \cr
DC:\quad &<1,2><a,b>; \ <1,2><b,c>; \ <1,2><a,c> \cr
E:\quad &<3,a><b,c>; \ <3,a,b>; \ <3,a,c>  \cr
EC:\quad &<1,2><3,a>; \ <1,2><3,a,b,c>; \ <1,2><3,a,c,b>\ , \cr} \eqn\DEassign$$

The first of these identifications can actually be verified, because
we can compute the NIMrep matrices for the simple current extension
of $A_2$ level 9 using the formalism developed in \FHSSW. Consider the
matrices $A^{i~~b}_a$, with $i$ a zero charge primary. Since the boundaries
are in one-to-one correspondence with simple current orbits, we can assign
the corresponding charge to each boundary. As is well known \FuchsXN,
zero charge
boundaries are symmetry preserving, the others symmetry breaking.
Now consider the matrix elements $A^{i_0~~b_0}_{~a_0}$,
$A^{i_0~~b_1}_{~a_1}$, $A^{i_0~~b_2}_{~a_2}$, where the subscript on the
labels denotes three times the charge. As expected, these matrix elements
are precisely given, respectively, by the NIMreps for the modular invariants
$<>$, $<a,b,c>$ and $<a,c,b>$ (we assume that the two mutually transpose
solutions in row 11 of the table are associated with the latter two
modular invariants).
 All other matrix elements of $A^{i_0}$ vanish due to
charge conjugation. With considerably more work one should be able
to compute also the NIMreps for charged primaries. This requires
extracting the boundary coefficients from the NIMreps and in particular
resolving the phase ambiguities.

The same comparison made above for the $D$ invariant, can in principle
also be made for the other three invariants, $DC, E$ and $EC$ of $A_2$ level
9. This requires an explicit expression for the corresponding NIMreps of
the unextended theory. These NIMreps can in principle be extracted
with a considerable amount of work from
\DiF\ZuberIA\ (the four cases D, DC, E and EC correspond to
${\cal D}^{(12)*}$, ${\cal D}^{(12)}$,  ${{\cal E}_4}^{(12)*}$ and
${{\cal E}_5}^{(12)}$ in the notation of these authors; the
remaining cases are discussed at the end of this section).

Remarkably the following four cases do not work (we list here
symmetric invariants and their cyclic permutations)
$$ D: \quad <1,2>; \ <1,2><a,b,c>; \ <1,2><a,c,b>  $$
This is the diagonal invariant of the extension. None of the three permutations
admits any NIMreps.
$$ DC: \quad <a,b>; \ <b,c>; \ <a,c>  $$
Again, none of the permutations has a NIMrep.
$$ E: \quad <3,a>; \ <3,a,b,c>; \ <3,a,c,b>  $$
Here it seems that the two permutations of $<3,a>$
do admit NIMreps, but that solution
just happens to coincide with the one of $<3,a><b,c>$. Apparently this  
solution should be assigned
to the latter modular invariant, which is needed in \DEassign. This also  
explains why it
exists on non-orientable surfaces, not admitted by the modular invariant $<3,a,b,c>$.
$$ EC: \quad <1,2><3,a><b,c>; \ <1,2><3,a,b>; \ <1,2><3,a,c>  $$
This is precisely the opposite of the previous case: there appears to be a
solution for $<1,2><3,a><b,c>$, but it does not admit a crosscap coefficient,
and should be assigned to $<1,2><3,a,b,c>$, needed in \DEassign. We see thus  
that there is precisely
one way of relating the four modular invariants of the unextended theory to
groups of three modular invariants of the extended theory, related by $Z_3$
permutations. The solutions 8 and 9 are fake ones. Removing them we get also
a consistent result for the absence of solutions for $Z_3$-related modular  
invariants.

\subsection{The extension to $E_6$}

There is an additional
modular invariant of $A_2$ level 9 corresponding to the conformal embedding in 
$E_6$. It can be obtained by means of an exceptional extension on top of the
simple current extension. The case has the interesting feature
of allowing two (or three) distinct NIMreps (see refs. \DiF\ZuberIA).
$$ |[1]+[4]|^2+2 \times |[5]|^2 \eqn\invesix$$
We have analyzed this modular invariant from the point of view of the
extended algebra, and find a result that seems in agreement with
\DiF\ZuberIA: in a complete search we did indeed find precisely three distinct  
NIMreps.
Each of these has
4 S-NIMreps, but only two of the three NIMreps allow a U-NIMrep. These
two distinct U-NIMreps differ in a very interesting way: one of them
has a Klein bottle coefficient equal to 0 on the degenerate field
(\ie\ the field [5]),
the other has this coefficient equal to 2. The interpretation is now
immediately obvious. The extended theory, $E_6$ level 1, is complex, and
has two modular invariants, charge conjugation and the diagonal invariant.
Both invariants have a U-NIMrep, according to refs.
\Cardy\ and \FSB\ respectively.
The Klein bottle coefficients are respectively $(1,0,0)$ and $(1,1,1)$.
In terms of the exceptional invariant \invesix, which cannot distinguish
the complex field from its conjugate, these two cases reveal themselves
through a different value of the Klein bottle coefficient on the degenerate
field, just as discussed in \HuiszoonGE.
The main novelty in this example is the fact that the two distinct
Klein bottle choices (which we knew {\it a priori}) leads to two
distinct NIMreps, whereas in all cases studied so far (\ie\ the simple
current extensions) different Klein bottle choices gave rise to different
unoriented annuli, but the same NIMrep.

Note that the existence of two distinct NIMreps seems to clash with
the discussion on uniqueness of the classifying algebra (and
hence its representations) in chapter 2.
This example suggests a very
obvious way out. Clearly the modular invariant \invesix\ belongs
to two distinct CFT's. Hence it seems plausible that after a proper
resolution of the degeneracy of the field $[5]$ one will in fact
obtain two distinct sets of classifying algebra coefficients, although
the details will have to be worked out. This phenomenon can be expected
to occur in general for modular invariants with multiplicities larger than
1, except for simple current invariants.

The two NIMreps that admit a U-NIMrep probably correspond to the
cases ${\cal E}_1^{(12)}$ and ${\cal E}_2^{(12)}$ in \ZuberIA, whereas
the third NIMrep is likely to correspond to ${\cal E}_3^{(12)}$, which
is discarded in \ZuberIA\ for not very transparent reasons.

Since the proper interpretation of these NIMreps appears to be a longstanding
problem in the literature \DiF\ZuberIA\GannonKI\ we present the
NIMreps here explicitly.
We present them here  only in terms of the simple current
extension of $A_2$ level 9\rlap.\foot{We
have also obtained the three solutions in the unextended theory, where
for this modular invariant an exhaustive search was possible. However
in the unextended theory one gets
55 $12 \times 12$ matrices, which we will not present here.}

The one with Klein bottle coefficients $K^0=K^4=1$, all others zero is:

$$ A^0 =\pmatrix{       1 & 0 & 0 & 0 \cr
		  0 & 1 & 0 & 0 \cr
		  0 & 0 & 1 & 0 \cr
		  0 & 0 & 0 & 1 \cr} \ \ \ \
A^4 =\pmatrix{       1 & 0 & 0 & 2 \cr
		  0 & 1 & 0 & 2 \cr
		  0 & 0 & 1 & 2 \cr
		  2 & 2 & 2 & 13 \cr}  \ \ \ \
A^5 =\pmatrix{      0 & 1 & 1 & 2 \cr
		  1 & 0 & 1 & 2 \cr
		  1 & 1 & 0 & 2 \cr
		  2 & 2 & 2 & 14 \cr}   $$
$$ A^i =\pmatrix{       0 & 0 & 0 & 1 \cr
		  0 & 0 & 0 & 1 \cr
		  0 & 0 & 0 & 1 \cr
		  1 & 1 & 1 & 6 \cr} \ ,\ \ \  i=1,2,3,6^1,6^2,6^3 $$
The one with Klein bottle coefficients $K^0=K^4=1$, $K^5=2$, all others zero is: 

$$ A^0 =\pmatrix{       1 & 0 & 0 & 0 \cr
		  0 & 1 & 0 & 0 \cr
		  0 & 0 & 1 & 0 \cr
		  0 & 0 & 0 & 1 \cr} \ \ \ \
A^4 =\pmatrix{       1 & 2 & 2& 2 \cr
		  2 & 5 & 4 & 4 \cr
		  2 & 4 & 5 & 4 \cr
		  2 & 5 & 5 & 4 \cr}  \ \ \ \
A^5 =\pmatrix{      2 & 2 & 2 & 2 \cr
		  2 & 4 & 5 & 5 \cr
		  2 & 5 & 4 & 5 \cr
		  2 & 5 & 5 & 4 \cr}   $$
$$ A^i =\pmatrix{       0 & 1 & 1 & 1 \cr
		  1 & 2 & 2 & 2 \cr
		  1 & 2 & 2 & 2 \cr
		  1 & 2 & 2 & 2 \cr} \ ,\ \ \  i=1,2,3,6^1,6^2,6^3 $$
and finally the one not admitting any Moebius coefficients is
$$ A^0 =\pmatrix{       1 & 0 & 0 & 0 \cr
		  0 & 1 & 0 & 0 \cr
		  0 & 0 & 1 & 0 \cr
		  0 & 0 & 0 & 1 \cr} \ \ \ \
A^4 =\pmatrix{       3 & 2 & 2& 4 \cr
		  2 & 3 & 2 & 4 \cr
		  2 & 2 & 3 & 4 \cr
		  4 & 4 & 4 & 7 \cr}  \ \ \ \
A^5 =\pmatrix{      2 & 3 & 3 & 4 \cr
		  3 & 2 & 3 & 4 \cr
		  3 & 3 & 2 & 4 \cr
		  4 & 4 & 4 & 8 \cr}   $$
$$ A^i =\pmatrix{       1 & 1 & 1 & 2 \cr
		  1 & 1 & 1 & 2 \cr
		  1 & 1 & 1 & 2 \cr
		  2 & 2 & 2 & 3 \cr} \ ,\ \ \  i=1,2,3,6^1,6^2,6^3 $$

Looking at these matrices it is immediately clear that our interpretation
is indeed the correct one. Namely, the annulus partition function
in the first case, $\sum_i A^{i~~b}_{~a}\chi_i$ restricted to the
first three boundaries turns into
$$ (\chi_0+\chi_4)\delta_a^b + A^{5~~b}_a \chi_5 $$
where $(\chi_0+\chi_4)$ is the identity character of $E_6$ and
$\chi_5$ is the Virasoro character of the $(27)$ and $(\overline{27})$
representations (this follows directly from the conformal embedding).
By inspection, $A^{5~~b}_{~a}$ restricted to the first three boundaries
is the sum of the $E_6$ annulus coefficients for the
$(27)$ and the $(\overline{27})$. The fourth boundary must then be
a symmetry breaking one, that respects the simple current extension
of $A_2$ level 9, but not the exceptional extension on top of it.

The interpretation of the second NIMrep is
completely analogous, but less instructive,
because it only concerns the first boundary. The other three are
symmetry breaking. The last NIMrep does not admit a rewriting
in terms of $E_6$ characters for any choice of boundaries.

The power of orientation is of course very well demonstrated in the
latter example. We consider some other examples of this type in chapter 7.

\section{$A_4$ level 5}

Consider now the simple current extension
of $A_4$ level 5. This extension has 10 representations,namely
$$  (0): (0,0,0,0) ; \ (1): (1,0,0,1) ; \ (2): (2,0,0,2) $$
$$  (3): (0,2,2,0) ; \ (4): (0,1,1,0) ; \ (5^i): (1,1,1,1), i=1,\ldots,5 $$
All 720 permutations of the representations $(1)$ and $(5^i)$ yield
modular invariant partition functions. In this  case the
table of solutions (which is complete, as in the previous case) is as follows

\vskip .7truecm
\begintable
 Ish. |  NIMreps | S-NIMreps|  U-NIMreps \cr
    10        |        1        |     1          |      1   \nr
     8        |     0          |    0            |     0 \nr
   7        |     0          |    0            |     0  \nr
         6        |     5          |   $(1+1+1+1+2)$   | $(0+0+0+0+(0+2))$   \nr
    5        |     2          |    $(1+1)$            |   0    \nr
    4        |     1          |     1            |    0     \endtable

Since the six primaries $(1)$ and $(5^i)$
are completely on equal footing as far as the modular data
(\ie\ $S$ and $T$) are concerned,
the solution depends only on the number of Ishibashi states, i.e. the
number of primaries not involved in a permutation. The 720 modular
invariants are related to one of the following two modular
invariants of $A_4$ level 5
$$\eqalign{ D:\quad  &[0]^2 + [1]^2 + [2]^2 + [3]^2 + [4]^2 + 5 [5]^2 \cr
E:\quad  &[0]^2 + [1][5] + [5][1] + [2]^2 + [3]^2 + [4]^2 + 4 [5]^2 \ ,\cr} $$
where $[0]\ldots[4]$ denotes a complete simple current orbit. If $[1]$ is
non-trivially involved in the permutation, the invariant of the extended
theory is related to E, and otherwise to D.
The notation is as in the
previous table: for example, the permutation $<1,a><b,c>$ gives rise
to six Ishibashi states. We read from the table that there exist 5 NIMreps,
four of which yield one S-NIMrep, while the fifth one yields two. Of those
two, one allows no U-NIMreps, and one allows two (which are only very
marginally different: the Klein bottle coefficients are the same, but
a few Moebius coefficients have opposite signs). The five NIMreps
are all completely different, and clearly not related by some
unidentified symmetry.

The most striking feature is again the complete absence of NIMreps for
the symmetric automorphism modular invariants $<1,a>$ and $<a,b>$, which
presumably are unphysical. For the automorphism $<1,a><b,c><d,e>$ a
NIMrep and even an S-NIMrep is available, but it does not allow a
U-NIMrep.

As in the previous case, we can compare the NIMreps for the simple
current extension of $A_4$ level 5 with the results from the table.
This time we consider $A^{i_0~~b_0}_{~a_0}$,
$A^{i_0~~b_1}_{~a_1}$,
$A^{i_0~~b_2}_{~a_2}$,
$A^{i_0~~b_3}_{~a_3}$ and
$A^{i_0~~b_4}_{~a_4}$, and we find that the result coincides with
the NIMreps of $<>$ and $<a,b,c,d,e>$ in the table. In the latter case
(five Ishibashi states) we found two NIMreps, which coincide respectively
with the entries of the charge $\pm 1/5$ and $\pm 2/5$ boundaries.

Identifying other solutions is harder than in the previous case, partly
because there is an ordering problem in determining cyclic symmetries.
From the point of view of the fusion algebra, the five resolved
fixed point representations are completely on equal footing, so that
the fusion rules are invariant under all their permutations.
If we make the assumption that boundaries with broken symmetries
in the unextended case correspond to boundaries for cyclically
permuted modular invariants in the extended case, (as is the case
thus far), then an apparent contradiction
arises. For example, the two modular invariants $<1,5^1><5^2,5^3>$
and $<1,5^1><5^2,5^5>$, which appear to be totally equivalent,
yield a different structure of permutations if one cyclically
permutes the fixed point labels. The cyclic permutations of the
symmetric modular invariants are as follows
\vskip .7truecm
\begintable
Nr. | invariant | type | permutations  \cr
1 | $<ab>$ | D | $2<abcd>+2<abc><de>$ \cr
2 | $<1a>$ | E | $4<1abcde>$ \cr
3 | $<ab><cd>$ | D | $4<ab><cd>$ \cr
4 | $<1a><cd>$ | E | $2<1abcd>+2<1abc><de>$ \cr
5 | $<1a><cd>$ | E | $2<1abcd>+2<1ab><cde> $\cr
6 | $<1a><cd><ef>$ | E | $4<1ab><de>$ \cr
7 | $<1a><cd><ef>$ | E | $2<1abcde>+2<1abc> $\endtable

Here only the permutation group element is indicated.
where
$a,b,c,d,e$ stand for the fixed point fields
in an unspecified order. For example $<abc><de>$ indicates
a product of and order 3 and an order 2 cyclic permutation,
but is not meant to imply an actual ordering of the labels. In the last
four rows, the actual choice of labels in column 1 determines which of
the two options is valid. Note that the number of Ishibashi states
in columns 1 and 3 correctly adds up to the number of Ishibashi states
of the type D and E modular invariants (30 and 24, respectively).

Clearly, there is no NIMrep for cases 1 and 2, since there is
no NIMrep for the symmetric invariant. Presumably the same is true
for cases 6 and 7, since there is no U-NIMreps for the symmetric invariant,
which has 4 Ishibashi states. There does exist a NIMrep in this case,
but this is precisely needed for the a-symmetric invariants
$<1abc><de>$ and $<1ab><cde>$ appearing in case 4 and 5, which is
the only remaining chance
to realize the E-invariant in terms of the extended theory.

\chapter{Examples II: Diagonal invariants }

The problems that we encountered above with the diagonal invariant
of the extension of $A_2$ level 9 can be expected to occur quite
generally for similar invariants. Consider a CFT with a simple
current of odd (and, for simplicity, prime) order $N$ of integer spin. Suppose the
unextended CFT has a complete U-NIMrep for the diagonal invariant.
This assumption holds in particular for (tensor products of)
WZW models, for which the boundary states and NIMreps were constructed
by means of an charge conjugation orbifold theory \FSB. Although
orientation issues were not discussed in that paper, the results were
obtained using a simple current extension of the orbifold theory, to which
the results of \FHSSW\ apply. Those results can then be used to derive
also the crosscap states.

If a full U-NIMrep is available for the diagonal invariant,
this implies that there exists a set of Klein bottle coefficients
that satisfies the following sum rule
$$ \sum_j S_{ij}K^j = 0\ \ \  \hbox{~if $i$ is complex} $$
In the case of WZW models such a sum rule must in fact hold for $K^i=1$
(and also for $K^i=\nu^i$, the Frobenius-Schur indicator), for all $i$. It
should be possible to derive this directly
from the Kac-Peterson formula for $S$, but
follows in any case from the argument in the previous paragraph.

Now consider a complex representation $i$ with vanishing simple
current charge (so that it survives in the extended
theory), and whose conjugate lies on a different orbit.
Then all other representations on its orbit are necessarily complex as
well, and hence
$$ 0 = \sum_j S_{Ji,j}K^j = \sum_j S_{ij} e^{2\pi i Q_J(j)} K^j $$
This implies that the sum must vanish for each value of the simple current
charge separately, and in particular
$$ \sum_{j, Q_J(j)=0} S_{i,j}K^j = 0 \eqn\extrule $$
This appears to be the sum rule needed for the simple current extended theory,
and indeed it is if there are no fixed points. But if there are fixed points,
their total contribution to the sum is enhanced by a factor $N$, so that the
correct sum rule for the extended theory does not hold\rlap.\foot{It turns out
the the fixed point contribution is in general non-vanishing. For simplicity
we assume that the fixed point fields themselves are real.}

There are two ways to remedy this. One is to flip the signs of some
of the Klein bottle coefficients on the fixed points. This would violate
the Klein bottle constraint, but even if one accepts that, in the examples
we studied no corresponding complete set of boundaries could be found.
The second solution is to modify the modular invariant in such a way that
in addition to the charge conjugation, $N-1$ resolved fixed points are
off-diagonal. This removes $N-1$ contributions $K^j$ for each fixed point,
so that the sum rule holds again. Note that this also introduces
new sum rules for the new off-diagonally paired fields themselves, which are
no longer Ishibashi states. The latter sum rules automatically follow from
\extrule\ and the general structure of the matrix $S$ on resolved
fixed points \ScYe. Note, however, that for $N>3$ the $N-1$
off-diagonal elements
need not be paired: there are other (non-symmetric)
modular invariants in which they
are off-diagonal.

The foregoing argument only shows that a plausible set of Klein bottle
and crosscap coefficients can be found if in addition to the charge
conjugation one also pairs $N-1$ resolved fixed points off-diagonally,
simultaneously for all fixed points. It also shows that without the
latter, the existence of a U-NIMrep is unlikely. A few examples
of this kind can be studied.

In the series of simple current extension of $A_2$ level $3k$
the first complex theory is $A_2$ level 9, discussed in the previous
chapter. In the next case,  $A_2$ level 12, we found a U-NIMrep for
the diagonal invariant with interchange of two resolved fixed points, but no
NIMrep for the diagonal invariant or interchange invariant
separately (this statement is based on an exhaustive search).
Extrapolating these results to levels 3 and 6 (where the extended
theory is real) one might anticipate the existence of NIMreps
for the fixed point interchange invariants. This is indeed correct.
For $A_2$ level 3, the extension is $D_4$ level 1,
and the fixed point interchange is a simple current invariant,
hence the NIMrep
is the same as the one given in \FHSSW. Another example is the
extension of $E_6$ level 3 with the spin-2 simple current.  This has a novel
feature in having two resolved fixed points, $f^i$ and $g^i$,
$i=1,2,3$. Charge conjugation is trivial in the extended theory, and
the invariant $<f^1,f^2><g^1,g^2>$ has one NIMrep (with
one S- and U-NIMrep). Note that $<f^1,f^2><g^2,g^3>$ is not
modular invariant.

Other cases with similar features occur for tensor products and
coset CFT's. Examples are
$ A_{2,3} \times A_{2,3} $ and
$ A_{2,3} \times E_{6,3} $, extended
with the current $(J,J)$ and the coset CFT
 $ A_{2,3} \times A_{2,3} /A_{2,6}$, where the ``extension" is
by the identification current. In the first two cases, there is no NIMrep
for the diagonal invariant
and one NIMrep exists for the diagonal invariant plus fixed point
interchange (yielding 4 S-NIMreps and 2 U-NIMreps). This was
demonstrated using an exhaustive search. For the coset CFT
(and many analogous ones)
we
expect similar results,
but it was unfortunately not possible to verify this.

\chapter{Examples III: c=1 orbifolds}

Here we analyse some results from \NUNO\ on modular invariants of
$c=1$ orbifolds. In \NUNO\ we considered orbifolds at radius $R^2=2pq$
($p$, $q$ odd primes)
(with $pq+7$ primaries), and considered the automorphism that occurs
\FuchsSQ,\GannonJM\ if the radius factorizes into two primes.

Two cases where considered: the automorphism acting on the
charge conjugation invariant (denoted ``$C+A$" in \NUNO) and acting on the  
diagonal invariant (denoted ``$D+A$"). In both cases, two U-NIMreps
where found, but this was not claimed to be the complete set of solutions.
Meanwhile we have verified that it is indeed complete in the simplest case,
$pq=15$. The question whether the corresponding NIMreps are also distinct
was not addressed in \NUNO, but from the explicit boundary coefficients
one may verify that they are not. The solutions present some
features that are worth mentioning in the present context.

In the language of the present
paper the results can be described as follows

\item\dash $D+A$: one NIMrep, yielding 4 S-NIMreps. Of these four,
three do not allow any U-NIMreps and should presumably be viewed as
unphysical. The fourth one allows two distinct U-NIMreps, each with a different
Klein bottle as described in \NUNO.

\item\dash $C+A$: one NIMrep, yielding 2 S-NIMreps. Each of these
yields one U-NIMrep, with the two Klein bottles described in \NUNO.

Both cases contain a novelty with respect to the previous examples.
In the D+A case the two Klein bottle choices correspond to the same
S-NIMrep. Hence there is no change in the unprojected open string
partition function, but some of the Moebius signs are flipped. This
means that the only difference between the two orientation choices
is the symmetrization of the representations. With both choices, the
same branes are identified, but with different signs.

In the C+A case the novelty is that the boundary coefficients of the
two choices differ by 8th roots of unity rather than fourth roots, as
in all simple current cases.
This implies that at least one of the projections  $\Omega$ has
coefficients $\Omega^j$ that are phases, rather than signs.
This possibility is allowed by \Square\ provided that  
$\Omega^j=(\Omega^{j^c})^*$. We cannot check this directly because
we cannot determine $\Omega^j$, but we can at least check it for
the ratio of the two different orientifold projections.

From the
annuli $A^{\Omega_x}_{iab}$ we can compute $B_{ja}\Omega_x^j C^j B_{ja}$,
where $x=1,2$, labelling the two cases. Since neither $B$ nor $C$
depended on the orientifold projection, the ratio of these quantities
for $x=1,2$
determines the
ratio $\Omega_2^j/\Omega_1^j$. We find that this ratio is $\pm i$
on the twist fields,  $j=\sigma_i$ or $\tau_i$ (and $\pm 1$ on all other  
fields), and that it does
indeed respect \Square.

\chapter{Examples IV: Pure WZW automorphisms }

The automorphism modular invariants of WZW models were completely classified
in \GannonJM. The foregoing results on automorphisms of extended
WZW models may raise some doubts about their consistency as CFT's.
We will not answer that question completely here, but we will
consider the ``most exceptional" cases. The results are summarized
(together with those of the next section) in a table in the appendix.

Four basic types (which may be combined) may be distinguished:
Dynkin diagram automorphisms (\ie\ charge or spinor conjugation and
triality), simple current automorphisms, infinite series for SO(N) level
2 if $N$ contains two or more odd prime factors, and the three cases
$G_2$ level 4, $F_4$ level 3 and $E_8$ level 4. The first of these was
dealt with in \FSB, the second in \FHSSW, but combinations still have
to be considered. For even $N$, some cases follow from the results
for $c=1$ orbifolds, using the relation derived in \SchellekensWW, but
further work is needed for completing this and extending it to odd $N$.

This leaves the three "doubly exceptional" cases. The first two are
closely related since the fusing rings are isomorphic. The NIMrep of
$G_2$ level 4 is already known, and the one of $F_4$ level 3
turns out to be identical. In both cases we find four S-NIMreps, but
three of them do not allow a U-NIMrep. The fourth one gives rise to one
U-NIMrep.

For $E_8$ level 4 we find one NIMrep, yielding one S-NIMrep, which in
its turn gives one U-NIMrep. The matrix elements of
 $R_{ma}=B_{ma}\sqrt{S_{m0}}$  are all of the form
$\pm {2\over\sqrt{17}}\sin({2 \pi \ell\over 17})$, $\ell=1,\ldots 8$, and
are up to signs equal to those of the modular transformation matrix
of the twisted affine Lie algebra (denoted as in \FSSD)
$\tilde B^{(2)}_{1}$, level 14 or, equivalently,
 $\tilde B^{(2)}_{7}$, level 2, although
the significance of this observation is not clear to us.

\chapter{Examples V: WZW extensions }

In the table in the appendix
we list the number of (S-,U-)NIMreps
some extensions
of WZW-models. In all cases the results are based on a complete
search.
The extensions are either conformal embeddings
(denoted as $H \subset G$, listed in \CoSu),
simple current extensions (denoted as ``SC"),
or higher spin extensions (HSE).
If the CFT is complex, there are in principle two invariants, obtained
by extending the charge conjugation invariant or the diagonal invariant.
The latter possibility is indicated by a $(*)$ in the first column.
Most of the extensions of complex CFT's in the table are
themselves real (although
some become complex after fixed point resolution). In a few cases only
the extension of the diagonal invariant was tractable, since it has
fewer Ishibashi states than the extension of the charge conjugation invariant.

We only list those
simple current extensions where we found more
solutions than those obtained in \FHSSW, although we considered all
accessible low-level cases in order to check whether the set of solutions
of \FHSSW\ is complete. The higher spin extensions appeared in
the list of $c=24$ meromorphic CFT's \SchM,
although some were obtained before, or could be inferred from rank-level
duality. Since the existence of these CFT's is on less firm
ground than the existence of the other kinds of extensions, it is
especially interesting to find out if NIMreps exist.

In the U-NIMreps column those
cases denoted with a single asterisk have Klein bottle coefficients
violating the Klein bottle constraint
(but no other conditions); those with a double asterisk violate
the orientifold condition
$M^i_a = M^{i^c}_{a^c} $ (and often also
the Klein bottle constraint). The ones with a triple asterisk have
a value for $K_0$ equal to $-1$ (and hence violate the Klein
bottle constraint), and they also
violate the trace identity \kleintrace. Often these solutions also
violate the orientifold condition, and
in some case they give rise to imaginary crosscap coefficients.
We will assume that the latter violations are
unacceptable, but on the other hand we will be forced to conclude that
the Klein bottle constraint violations must be accepted.

All the U-NIMreps that are listed have
different Klein bottle coefficients. Those of the spin-1 extensions
can all be understood in terms of simple-current related Klein bottle
choices in the extended algebra  (which is a level-1 WZW model),
as described in \FHSSW.

Some of the NIMreps in
the table (in particular
those for the algebras of type $A$) have
been discussed elsewhere (see \eg\ \DiF\PetkovaFW), most others are
new, as far as we know. Although it is impossible to present all matrices
here explicitly, they are available on request.

The following features are noteworthy (the characters refer to the
last column)
\item{A.} In this case there are two NIMreps, one corresponding the interpretation
of the MIPF as a C-diagonal invariant of the extended algebra, and one
corresponding to the diagonal interpretation.

\item{B.} In this case there are two U-NIMreps,
related to the fact
that the extended algebra admits two Klein bottle choices, generated by
simple currents as discussed in \HSSK.

\item{C.} This is as case B, except that the second Klein bottle choice
violates the Klein bottle constraint in the unextended theory (although it
does satisfy it in the extended theory).

\item{D.} These are combinations of cases A, B and C: the modular invariant
corresponds to both diagonal and C-diagonal interpretations, each of which allows
two Klein bottle choices. Since the diagonal invariant of $D_{2n+1}$
is a simple current invariant, all four Klein bottle choices follow in fact
from \FHSSW.

\item{E.} In this case the extension is by a simple current, so that
 \FHSSW\ applies. However, that paper contains only a single NIMrep for this
case, and we find two. A plausible explication is that the $A_{2,3m}$
series bifurcates for $m \geq 3$ into a series of simple current modifications
of the C-diagonal and of the diagonal invariant of $A_{2,3m}$. For $m<3$ these
invariants coincide. The formula of \FHSSW\ only applies to the first series,
and the second solution must be interpreted as part of the other series. This
interpretation agrees with the values of the Klein bottle coefficients on the
fixed point field (resp. 3 and 1), and also with the discussion
at the end of section 4.

\item{F.} Here the same remarks apply as in case E, but in addition the
resolved fixed points become simple currents (of $SO(8)$ level 1),
allowing an additional Klein
bottle choice (or rather three, related by triality).
The three cases in the table (not including the double asterisk ones) have  
Klein bottle
values $3,1$ and $-1$ on the fixed point field. This corresponds precisely
to the three values that follow from \FHSSW\ for $SO(8)$ level 1 (for
respectively the diagonal invariant, the simple current automorphism and
the Klein bottle simple current).

\item{G.} Here ``SC 2" means that the invariant is obtained using
the simple current with Dynkin labels $J^2=(0,2,0)$. This case does not really
belong in the table, because both from the point of view of the
unextended theory, $A_3$ level 2, as in the extended theory, $A_5$
level 1, the entire result can be obtained using \FHSSW. In
$A_{3}$ level 2 they follow from
by combining both choices for
 $\alpha(J)$ in formula (11) of \FHSSW, applied to the extension by $J^2$,
 with the two allowed
Klein bottle choices, $K=1$ and $K=J$. In $A_5$ level 1 one obtains
the same four Klein bottles from two distinct modular invariants which
cannot be distinguished prior to fixed point resolution. The invariants are
the diagonal one, and the one generated by the simple current $J^2$
of $A_5$. Both can be
modified by the Klein bottle simple current J.

\item{H.} This is the unextended analog of the example that was
explicitly presented in section (3.2.1). It is in complete agreement
with the results presented there.

One can make sense of the higher spin extension results in a very similar
way, because also in those cases we know the fusion rules of the
extended algebras. In some cases this knowledge is available via
rank-level duality of the
unextended algebras. Some of these dualities are quite clear in the table,
for example $C_{7,2}$ with $C_{2,7}\equiv B_{2,7}$.
In other cases one can make use of the fact
that the exceptionally extended CFT's combine with known CFT's to form a meromorphic
CFT, and hence must have the complex conjugate matrix $S$ of that CFT.
For example, the $E_7$ level 3 extension has the same fusion rules
as $A_5$ level 1. The results for the latter CFT match those of the
exceptional extension: we find the same number of distinct U-NIMreps
as in case G. The same is true, with
appropriate modifications, in all other cases.

These examples demonstrate once more the importance
of orientation, as well as the completeness of the
results of \FHSSW. They also give the clearest evidence so
far that the Klein bottle constraint may well be correct in some form,
but that the precise formulation requires more work.

\chapter{Conclusions}

Orientation effects add useful additional structure to the
a NIMreps. Furthermore, given a NIMrep, this information
is relatively easy to get, and it is therefore a pity that
many authors decide to ignore it.

Taken into account orientation leads naturally to a different
formula for the annulus than the one commonly used in the NIMrep literature.
 The
distinctions between the expression
for oriented annuli, \ABBtwo\ and unoriented ones \AomegaTwo\ are:
\item\dash The former does not depend on the choice of orientifold
projection, the latter does;
\item\dash The former does not imply any boundary conjugation, the
latter does.

We have given examples of modular invariants without NIMreps (including,
rather surprisingly, diagonal invariants), modular invariants with
several NIMreps,
NIMreps without
S-NIMreps (and hence no boundary conjugation), NIMreps with
several S-NIMreps, S-NIMreps without U-NIMreps and S-NIMreps with
more than one U-NIMrep, and U-NIMreps with
identical Klein bottle coefficients but different Moebius coefficients.

Two distinct modular invariants may have the same Ishibashi states and
hence the same NIMreps. This happens in particular when asymmetric
invariants are considered. In that case the existence of a U-NIMrep is
useful information to determine which NIMrep belongs to which
modular invariant.

Although there  appear to exist NIMreps that are accidental solutions
to the equations, and that have no physical interpretation, we
have not found any U-NIMreps that are clearly not sensible.

In all cases we know, the occurrence of multiple NIMreps
for symmetric modular invariants can be interpreted
in a sensible way if orientation information is taken into account, and if
those NIMreps that lack a corresponding U-NIMrep are rejected.
Furthermore, the occurrence of multiple U-NIMreps is in complete agreement
with the results of \FHSSW, wherever a comparison is possible.
It seems plausible that only if the partition function has multiplicities
larger than 1, there can be more than one ``sensible" NIMrep.

The results in section 7 (including a large set of cases not appearing
in the table because they agree with expectations)
suggest that the results of \FHSSW\ are in fact generically complete, \ie\
no other boundary and crosscap states can be written down that are valid
for any simple current invariant. Sporadically there may be additional
solutions, but the absence of these solutions in other cases with the
same simple current orbit structure proves that they are not
generic (unfortunately
there is a infinity of distinct orbits structures, and hence ``generic
completeness" cannot be proved rigorously in this way).

Although our results shed some light on the infamous "Klein bottle constraint",
its precise formulation clearly requires
further thought. The examples in the last section show
that it does not hold in its naive form, at least not for
modular invariants of extension type.

The purpose of the examples in this paper is only to demonstrate various
features. They should be useful to avoid incorrect conjectures, but no
general conclusion can be drawn from them with confidence.
Indeed, even the combined information from the existence of a modular
invariant, NIMreps, S-NIMreps and U-NIMreps is probably
ultimately insufficient
to decide whether a given candidate CFT really exists. Fortunately,
in applications to string theory we rarely need to consider such
exotic cases. Simple current invariants is all one usually needs, and
for that class the solution is known.

\ack

We would like to thank  L. Huiszoon, Ya. Stanev, A. Sagnotti, C. Schweigert,
and I. Runkel for discussions. Partial checks on the results of
section 3 have been done by Ya. Stanev. Special thanks to C. Schweigert
and L. Huiszoon for constructive comments on the manuscript.
A.N. Schellekens wishes to thank the theory group of Tor Vergata, where part  
of this work was done,
for hospitality and the INFN for financial support.
N.Sousa wishes to thank Stichting F.O.M. and Funda\c c\~ao para a Ci\^encia
e Tecnologia for financial support under the reference BD/13770/97.

\par \penalty-4000\vskip\chapterskip
   \spacecheck\referenceminspace \immediate\closeout\referencewrite
   \referenceopenfalse
   \line{\fourteenrm\hfil REFERENCES\hfil}\vskip\headskip
   \endlinechar=-1
   \input referenc.texauxil
   \endlinechar=13
   
\endpage
\appendix{Number of NIMreps for WZW modular invariants}
\vskip .7truecm
\begintable
MIPF                   | NIM | S-NIM | U-NIM |  remark \cr
$A_{1,10} \subset SO(5)$ | 1 | 2 | $1\!+\!1$ | B \nr
$A_{1,16} \hbox{(``$E_7$" inv.)}$ | 1 | 1 | 1 | \nr
$A_{1,28} \subset G_2$ | 1 | 1 | 1 | \nr
$A_{2,3}$ (SC/$\subset SO(8)$) | 2 | 2+3 |  
$(0\!+\!1)\!+\!(0\!+\!1\!+\!(1^*\!\!+\!1^{***}))$| F \nr
$A_{2,5} \subset SU(6)$ | 1 | 2 | $1\!+\!1^*$| C \nr
$A_{2,6}$ (SC) | 2 | 2+1 | $(0\!+\!1)\!+\!(1)$| E \nr
$A_{2,9} \subset E_6$ | 3 | 2+2+2 |  
$(1\!+\!0\!)\!+\!(1\!+\!0\!)\!+\!(\!0\!+\!0)$| H \nr
$A_{3,2}$ (SC 2/$\subset SU(6)$) | 1 | 4 | $1\!+\!1\!+\!1\!+\!1$
| G \nr
$A_{3,4} \subset SO(15)$ | 1 | 4 | $0\!+\!0+\!(1+2^{**})\!+\!(1+2^{**})$ | B \nr
$A_{9,2}$ (HSE$^*$)| 1 | 2 | $1+\!1$ | \nr
$B_{2,3} \subset SO(10)$ | 2 | 4+4 |  
$(0\!+\!0\!+\!1\!+\!(\!1^*\!\!+\!1^{***}))\!+\!(0\!+\!0\!+\!1\!+\!1)$ | D \nr
$B_{2,7} \subset SO(14)$ | 2 | 4+4 |  
$(0\!+\!0\!+\!1\!+\!(1^*\!\!+\!1^{***}))\!+\!(0\!+\!0\!+\!1\!+\!1)$ | D \nr
$B_{2,12} \subset E_8$ | 1 | 4 | 1 | \nr
$B_{12,2}$ (HSE) | 1 | 2 | 1 | \nr
$C_{3,2} \subset SO(14)$ | 2 | 4+4 |  
$(0\!+\!0\!+\!1+\!(1^*\!\!+\!1^{***}))\!+\!(0\!+\!0\!+\!1\!+\!1)$ | D \nr
$C_{3,4} \subset SO(21)$ | 1 | 16 | $1\!+\!1\!+\!14 \! \times\! 0$ | B \nr
$C_{4,3} \subset SO(27)$ | 1 | 16 | $1\!+\!1\!+\!14 \! \times\! 0$ | B \nr
$C_{7,2}$ (HSE) | 2 | 4+4 |  
$(0\!+\!0\!+\!1\!+\!(1^*\!\!+\!1^{***}))\!+\!(0\!+\!0\!+\!1\!+\!1)$ | \nr
$D_{7,3}$ (HSE$^*$) | 2 | 2+2 | $(\!0\!+\!0\!)\!+\!(\!0\!+\!2\!)$ | \nr
$D_{7,3}$ (HSE) | 2 | 2+2 | $(\!0\!+\!0\!)\!+\!(\!1\!+\!1\!)$ | \nr
$C_{10,1}$ (HSE) | 1 | 2 | $1+\!1$ | \nr
$D_{9,2}$ (HSE) | 2 | 2+5 |
$(\!1\!+\!1\!)\!+\!(5\times 0)$ | \nr
$D_{9,2}$ (HSE$^*$) | 3 | 4+4+4 |
$(\!1\!+\!1\!+\!0\!+\!0)\!+\!(\!4 \!\times\! 0)\! +\!(\!4\! \times\! 0\!)$ | \nr
$G_{2,3} \subset E_6$ | 2 | 2+2 | $(0\!+\!1)\!+\!(0\!+\!1)$ | A \nr
$G_{2,4}$ (automorphism) | 1 | 4 | 1 | \nr
$G_{2,4} \subset SO(14)$ | 2 | 4+4 |  
$(0\!+\!0\!+\!1\!+\!(\!1^*\!\!+\!1^{***}))\!+\!(0\!+\!0\!+\!1\!+\!1^*)$ | D \nr
$F_{4,3}$ (automorphism) | 1 | 4 | 1 | \nr
$F_{4,3} \subset SO(26)$ | 2 | 4+4 |  
$(0\!+\!0\!+\!1\!+\!(\!1^*\!\!+\!1^{***}))\!+\!(0\!+\!0\!+\!1\!+\!1^*)$ | D \nr
$E_{6,4}$ (HSE$^*$) | 2 | 2+2 | $(\!0\!+\!0)\!+\!(1\!+\!1^*)$ | \nr
$E_{7,3}$ (HSE) | 2 | 4+4 | $2 \times (\!0\!+\!0\!+\!1\!+\!1)$ | \nr
$E_{8,4}$ (automorphism) | 1 | 1 | 1 |
\endtable
\vskip 1.truecm

\end